\documentclass[twocolumn,byrevtex,prb,showpacs]{revtex4}

\usepackage{graphicx}
\usepackage{amsmath2000}   
 
\begin{document}

\title{Dynamical density-matrix renormalization-group method} 

\author{Eric Jeckelmann}
\affiliation{Fachbereich Physik, Philipps-Universit\"{a}t, 
D-35032 Marburg, Germany}

\date{\today}

\begin{abstract}
I present a density-matrix renormalization-group (DMRG) 
method for calculating dynamical properties and excited states 
in low-dimensional lattice quantum many-body systems. 
The method is based on an exact variational principle for dynamical
correlation functions and the excited states contributing to them.
This dynamical DMRG is an alternate formulation of the 
correction vector DMRG but is both simpler and more accurate.
The finite-size scaling of spectral functions is discussed and a method
for analyzing the scaling of dense spectra is described. 
The key idea of the method is a size-dependent broadening of the
spectrum.
The dynamical DMRG and the finite-size scaling analysis are 
demonstrated on the optical conductivity of the one-dimensional
Peierls-Hubbard model.
Comparisons with analytical results show that the spectral functions
of infinite systems can be reproduced almost exactly with these 
techniques.
The optical conductivity of the Mott-Peierls insulator is investigated
and it is shown that its spectrum is qualitatively different from the
simple spectra observed in Peierls (band) insulators and one-dimensional
Mott-Hubbard insulators.
\end{abstract}

\pacs{71.10.Fd, 78.20.Bh}

\maketitle

\section{Introduction \label{sec:intro}}

The density-matrix renormalization group (DMRG)~\cite{steve,dmrgbook}
is a very successful numerical methods for calculating
static properties of ground states and low-lying eigenstates in quantum
many-body systems. 
For low-dimensional strongly correlated systems DMRG is as accurate
as exact diagonalization techniques but can be used to study much larger
systems than with those techniques (currently, up to $\sim 10^3$ sites).
Using a finite-size scaling analysis it is thus possible to determine 
the static properties of a system in the thermodynamic limit with great 
accuracy.

The calculation of dynamical properties and higher energy excitations
with DMRG has proved to be more difficult.
Several approaches have been proposed but calculations have been
carried out successfully for few problems only. 
The simplest of these methods is the Lanczos 
DMRG.~\cite{karen,till}
In practice, this method gives accurate results for the first few 
moments of a dynamical spectrum.
Therefore, it works well for simple discrete spectra
made of a few peaks but it usually fails for more complicated spectra.
An alternate method for calculating dynamical properties is the 
correction vector DMRG.~\cite{pati,till}
Contrary to the Lanczos DMRG, this method can describe 
complex or dense spectra accurately. 
Nevertheless, there have been few applications~\cite{till2,brune}
of correction vector DMRG until now because this method is more 
difficult and requires significantly more computer resources than 
the usual DMRG method for calculating static properties at low energy.

In this paper I describe a simple and efficient method, called
the dynamical DMRG (DDMRG), for calculating dynamical
properties and excited states with DMRG.
This approach is based on a variational principle
for dynamical correlation functions and the related excited states.
The variational principle is essentially an elegant
formulation of the correction vector technique.
Because of the variational formulation, however, the DDMRG method is 
easier to use and significantly more accurate than the correction 
vector DMRG method.

While the spectrum of a finite system is necessarily discrete,
continuous excitation bands are often found in the thermodynamic
limit. 
It is possible to broaden finite-size spectra to simulate the continuum 
of an infinite-system spectrum.
Usually, the broadening is arbitrarily large and no systematic
quantitative analysis of finite-size effects is performed
for the spectrum. 
Here I show that dynamical properties of infinite systems
can be obtained reliably using an appropriate finite-size scaling 
analysis.
The key to the analysis is the use of a broadening which scales 
systematically with the system size. 

The DDMRG method and the finite-size scaling technique for
dynamical spectrum have already been successfully used to investigate
the optical properties of one-dimensional Mott 
insulators.~\cite{eric,fabian}
Here I apply these techniques to the calculation of the
optical conductivity in the one-dimensional Peierls-Hubbard model
of conjugated polymers.~\cite{dionys}
Much effort has been devoted to understanding the optical properties
of these materials. 
In particular, the optical conductivity of the Peierls-Hubbard
model has been studied extensively and analytical results have 
been obtained  for various special cases, such as the Mott-Hubbard 
insulator and the Peierls (band) insulator limits.
Leaving out these special limits the Peierls-Hubbard model describes a 
Mott-Peierls insulator and its optical properties 
are still poorly understood.
In this paper I show that DDMRG can accurately reproduce the known 
analytical results for the optical spectrum in the thermodynamic limit. 
Then I investigate the optical conductivity of a Mott-Peierls
insulator using DDMRG and show that it displays specific features.

The paper is organized as follows:
The variational principle for dynamical correlation functions
and related excited states is presented in the next section.
I describe the dynamical DMRG method in Sec.~\ref{sec:ddmrg}. 
The finite-size scaling analysis is presented in Sec.~\ref{sec:scaling}.
I report and discuss the results for the optical conductivity of the 
Peierls-Hubbard model in Sec.~\ref{sec:model}.
Finally, I conclude in the last section.

\section{Variational principle \label{sec:principle}}

The dynamic response of a quantum system to a time-dependent
perturbation is often given by dynamical correlation functions 
such as
\begin{equation}
G_{A}(\omega + i \eta) = - \frac{1}{\pi} 
\langle \psi_0| A^{\dag} \frac{1}{E_0+\omega + i \eta - H} A 
|\psi_0\rangle \; ,
\label{dynamCF}
\end{equation}
where $H$ is the time-independent Hamiltonian of the
system, $E_0$ and $|\psi_0 \rangle$ are its ground-state energy and 
wavefunction, $A$ is the quantum operator corresponding to the physical
quantity which is analyzed, and $A^{\dag}$ is the Hermitian conjugate
of $A$. A small real number $\eta > 0$ is used in the calculation to 
shift the poles of the correlation function into the complex plane.  
(I set $\hbar=1$ in sections~\ref{sec:principle} to~\ref{sec:scaling}.) 

In general, we are interested in calculating the imaginary part
of the correlation function
\begin{eqnarray}
I_{A}(\omega + i\eta) &=& \text{Im} \ G_{A}(\omega + i \eta) \\
& = & \frac{1}{\pi} \langle \psi_0 | A^{\dag} 
\frac{\eta}{(E_0+\omega -H)^2 + \eta^2} A |\psi_0 \rangle \; \nonumber 
\end{eqnarray}
in the limit $\eta \rightarrow 0$
\begin{equation}
I_{A}(\omega) = \lim_{\eta \rightarrow 0} I_{A}(\omega + i \eta) 
\; \geq 0 \;.
\end{equation}
It should be noted that the spectrum $I_{A}(\omega + i \eta)$ for any
finite $\eta > 0$ can be calculated from the spectrum $I_{A}(\omega)$
by convolution with a Lorentzian distribution 
\begin{equation} 
I_{A}(\omega + i \eta) =  C_{\eta}[I_{A}(\omega)]  \; >  0 \; ,
\end{equation}
where I use the notation $C_{\eta}[f(\omega)]$ to represent the
convolution of a spectral function $f(\omega)$ 
with a Lorentzian distribution of width $\eta$
\begin{equation}
C_{\eta}[f(\omega)] = \int_{-\infty}^{+\infty} d\omega' f(\omega')
\frac{1}{\pi}\frac{\eta}{(\omega-\omega')^2+\eta^2} \; .
\end{equation}
The moments of the spectrum $I_{A}(\omega)$ fulfill sum rules
such as
\begin{eqnarray}
\int_{-\infty}^{+\infty} d\omega I_{A}(\omega)
& = & \langle \psi_0 | A^{\dag} A | \psi_0 \rangle \; , \nonumber \\
\int_{-\infty}^{+\infty} d\omega I_{A}(\omega) \omega
& = & \langle \psi_0 | A^{\dag} [H,A] | \psi_0 \rangle \; , 
\label{sumrules} \\
\int_{-\infty}^{+\infty} d\omega I_{A}(\omega) \omega^2
& = & \langle \psi_0 | [A^{\dag},H] [H,A] | \psi_0 \rangle \; , 
\nonumber
\end{eqnarray}
where  $[A,B] = AB-BA$.

A dynamical correlation function~(\ref{dynamCF}) can be calculated 
using the correction vector method.
The correction vector associated with $G_A(\omega + i \eta)$ is defined
by 
\begin{equation}
|\psi_A(\omega + i \eta) \rangle = \frac{1}{E_0+\omega + i \eta - H} 
| A \rangle \; ,
\end{equation}
where $| A \rangle = A | \psi _0 \rangle$.
If the correction vector is known, the dynamical correlation
function can be calculated directly
\begin{equation}
G_A(\omega + i \eta) = 
-\frac{1}{\pi} \langle A|\psi_A(\omega + i \eta) \rangle \; .
\label{dynamCF2}
\end{equation}
To calculate a correction vector one first solves 
an inhomogeneous linear equation 
\begin{equation}
\left [ (E_0+\omega-H)^2+\eta^2 \right ] | \psi \rangle
= -\eta | A \rangle \; ,
\label{CVequation1}
\end{equation}
which always has a unique solution 
$| \psi \rangle = | Y_A(\omega + i \eta) \rangle$ for $\eta \neq 0$.
The correction vector is then given by
\begin{equation}
|\psi_A(\omega + i \eta) \rangle = | X_A(\omega + i \eta) \rangle
+ i | Y_A(\omega + i \eta) \rangle  
\end{equation}
with
\begin{equation}
| X_A(\omega + i \eta) \rangle =  
\frac{H-E_0-\omega}{\eta} | Y_A(\omega + i \eta) \rangle  \; .
\label{CVequation2}
\end{equation}
One should note that the states $| X_A(\omega + i \eta) \rangle$
and $| Y_A(\omega + i \eta) \rangle$ are complex if the state 
$|A\rangle$ is not real, but they  always determine the real part and 
imaginary part of the dynamical correlation function 
$G_A(\omega + i \eta)$, respectively, 
\begin{subequations}
\begin{eqnarray}
\text{Re} \ G_A(\omega + i \eta) & =  &
-\frac{1}{\pi} \langle A|X_A(\omega + i \eta) \rangle \; ,
\label{ReDynamCF} 
\\
\text{Im} \ G_A(\omega + i \eta) & =  &
-\frac{1}{\pi} \langle A|Y_A(\omega + i \eta) \rangle \;  .
\label{ImDynamCF}
\end{eqnarray}
\end{subequations}
The derivatives of the real and imaginary parts can also be 
calculated from these states 
\begin{eqnarray}
\frac{d}{d\omega} \text{Re} G_A(\omega + i \eta) & =  &
\frac{1}{\pi} [ \langle X_A(\omega + i\eta)|X_A(\omega + i\eta) 
\rangle \nonumber \\
& &-\langle Y_A(\omega + i\eta)|Y_A(\omega + i\eta) \rangle] 
\label{derivatives}\\
\frac{d}{d\omega} \text{Im} G_A(\omega + i \eta) & =  &
\frac{2}{\pi} \langle X_A(\omega + i\eta)|Y_A(\omega + i\eta) \rangle
\; . \nonumber
\end{eqnarray}

A well-established approach for solving an inhomogeneous linear 
equation~(\ref{CVequation1}) is to formulate it as a minimization 
problem.
One considers the functional 
\begin{eqnarray}
W_{A,\eta}(\omega, \psi) & = &
\langle \psi | (E_0+\omega-H)^2+\eta^2  | \psi \rangle
\nonumber \\
& & 
+ \eta \langle A | \psi \rangle + \eta \langle \psi | A \rangle \; .
\label{functional}
\end{eqnarray}
For any $\eta \neq 0$ and a fixed frequency $\omega$ this functional 
has a well-defined and non-degenerate minimum 
for the quantum state which is solution of Eq.~(\ref{CVequation1})
\begin{equation}
| \psi_{\text{min}} \rangle = | Y_A(\omega + i \eta) \rangle \; .
\end{equation}

It is easy to show that the value of the minimum is related
to the imaginary part of the dynamical correlation function
\begin{equation}
W_{A,\eta}(\omega, \psi_{\text{min}}) =  
-\pi\eta I_A(\omega + i \eta).
\end{equation}
Therefore, the calculation of spectral functions can be formulated
as a minimization problem. 
To determine $I_A(\omega + i \eta)$ at any frequency $\omega$ and for 
any $\eta > 0$, one minimizes the corresponding functional 
$W_{A,\eta}(\omega, \psi)$.   
Once this minimization has been carried out, the real part of 
the correlation function $G_A(\omega + i \eta)$ can be calculated using
Eqs.~(\ref{CVequation2}) and (\ref{ReDynamCF}) if necessary.
This is the variational principle for dynamical correlation functions.

It is clear that if we can calculate $| Y_A(\omega + i \eta) \rangle$
exactly, this variational formulation is completely equivalent to the 
correction vector method.  
However, if we can only calculate an approximate solution with an error
of the order $\epsilon \ll 1$, $|\psi\rangle = | Y_A(\omega + i \eta) 
\rangle + \epsilon |\phi\rangle$ with $\langle \phi|\phi \rangle=1$, 
the variational formulation is more accurate.
In the correction vector method the error in the spectrum 
$I_A(\omega + i \eta)$ calculated
with Eq.~(\ref{ImDynamCF}) is also of the order of $\epsilon$.  
In the variational approach it is easy to show that the error
in the value of the minimum $W_{A,\eta}(\omega, \psi_{\text{min}})$, 
and thus in $I_A(\omega + i \eta)$, is of the order of $\epsilon^2$.
With both methods the error in the real part of $G_A(\omega + i \eta)$
is of the order of $\epsilon$. 

One can write the function $I_A(\omega)$ in the spectral form
(or Lehmann representation)
\begin{equation}
I_A(\omega) = \sum_{n} |\langle \psi_n|A|\psi_0\rangle|^2 
\delta(\omega + E_0-E_n)
\; ,
\end{equation}
where $|\psi_0 \rangle$ is the ground state, $|\psi_n\rangle, n > 1, $ 
denotes the other eigenstates of $H$, and $E_0, E_n$ are their 
respective eigenenergies. 
Obviously, only the eigenstates with a finite matrix element 
$\langle \psi_n|A|\psi_0\rangle \neq 0$ contribute to the spectrum and
here we are only interested in those excited states. 
In the correction vector method the excitation energies $E_n-E_0$ 
and the spectral weights $|\langle \psi_n|A|\psi_0\rangle|^2$
can be obtained from the poles of 
$G_A(\omega + i \eta)$.  
The corresponding wavefunctions $|\psi_n\rangle$ can be calculated
by taking the $\eta \rightarrow 0$ limit of the correction vectors
\begin{equation}
|\psi_n\rangle \propto 
\lim_{\eta \rightarrow 0} | Y_A(E_n-E_0 + i \eta) \rangle \;.
\end{equation}

The excited states contributing to $G_A(\omega + i \eta)$ 
correspond to the local maxima of the spectrum
$I_A(\omega + i \eta)$ for small enough $\eta > 0$.
Therefore, they can also be obtained by minimization of the functional 
$W_{A,\eta}(\omega,\psi)$ with respect to both $\omega$ and $\psi$. 
The local minima of $W_{A,\eta}(\omega,\psi)$ are given by the 
conditions
\begin{eqnarray}
\omega_{\text{min}} + E_0 & = & 
\frac{\langle \psi_{\text{min}} |  H | \psi_{\text{min}} \rangle}
{\langle \psi_{\text{min}}| \psi_{\text{min}} \rangle} \; , \nonumber \\
|\psi_{\text{min}}\rangle &=& | Y_A(\omega_{\text{min}} + i \eta) 
\rangle  \; .
\end{eqnarray}
In the limit $\eta \rightarrow 0$, $\omega_{\text{min}}+E_0$ tends
to the energy $E_n$ of an eigenstate with finite spectral weight,  
$|\psi_{\text{min}}\rangle$ is equal to the corresponding eigenstate
$|\psi_n\rangle$ up to a normalization constant, and 
$-W_{A,\eta}(\omega_{\text{min}},\psi_{\text{min}})$ tends to 
the spectral weight $|\langle \psi_n|A|\psi_0\rangle|^2$.
This is the variational principle for excited states contributing to
a dynamical correlation function $G_A(\omega +i\eta)$.

Again this variational formulation is completely equivalent to the 
correction vector method if $|Y_A(\omega +i\eta)\rangle$ can be 
calculated exactly. 
In an approximate calculation, however, errors in the eigenenergies and 
spectral weights are of the order of $\epsilon$ with the correction 
vector method, while they are of the order of $\epsilon^2$ with
the variational formulation.

\section{Dynamical DMRG \label{sec:ddmrg}}

DMRG is a numerical method for calculating the properties of lattice 
quantum many-body systems.  
It is described in detail in several publications (for instance, 
see Refs.~\onlinecite{steve} and~\onlinecite{dmrgbook}).
DMRG can be considered as a variational approach. 
The system energy
\begin{equation}
E(\psi)= \frac{\langle \psi |H|\psi \rangle}{\langle \psi |\psi \rangle}
\label{energyfunc}
\end{equation}
is minimized in a variational subspace (the DMRG basis) of the system 
Hilbert space to find the ground-state wavefunction $|\psi_0\rangle$ 
and energy $E_0 = E(\psi_0)$.
If the ground-state wavefunction is calculated with an error of 
the order of $\epsilon \ll 1$ (i.e., $|\psi\rangle = |\psi_0\rangle + 
\epsilon |\phi\rangle$ with $\langle \phi | \phi \rangle = 1$), 
the energy obtained is an upper bound to the exact result and the error
in the energy is of the order of $\epsilon^2$ (as in all variational 
approaches).
In principle, the DMRG energy error is proportional to the weight of 
the density-matrix eigenstates discarded in the renormalization 
procedure.
This discarded weight can be reduced by increasing the number $m$ of 
density-matrix eigenstates kept in the calculation, which corresponds 
to an increase of the variational subspace dimension.
Therefore, the energy error systematically decreases with increasing 
$m$ in a DMRG calculation.   

The DMRG procedure used to minimize the energy 
functional~(\ref{energyfunc}) can also be used to minimize the 
functional $W_{A,\eta}(\omega,\psi)$ and thus to calculate the 
dynamical correlation function $G_A(\omega+i\eta)$.
I call this approach the dynamical DMRG method. 
The minimization of the functional $W_{A,\eta}(\omega,\psi)$ is 
easily integrated into the usual DMRG algorithm.
At every step of a DMRG sweep through the system lattice,
a superblock representing the system is built and the following 
calculations are performed in the the superblock subspace: 
\begin{enumerate}

\item The energy functional $E(\psi)$ is minimized using a standard
iterative algorithm for the eigenvalue problem. This yields the ground 
state vector $|\psi_0\rangle$ and its energy $E_0$ in the superblock 
subspace.

\item The state $|A \rangle$ is calculated. 

\item The functional $W_{A,\eta}(\omega,\psi)$ is minimized using an
iterative minimization algorithm.
This gives the first part of the correction vector 
$|Y_{A}(\omega+ i\eta)\rangle$ and the imaginary part 
$I_A(\omega + i \eta)$ of the dynamical correlation function.

\item  The second part  $|X_{A}(\omega+ i\eta)\rangle$ 
of the correction vector is calculated using Eq.~(\ref{CVequation2}).

\item
The real part and the derivatives of the dynamical correlation function
can be calculated from Eqs.~(\ref{ReDynamCF}) and~(\ref{derivatives}),
respectively.

\item The four states $|\psi_0\rangle$, $|A \rangle$, 
$|Y_{A}(\omega+ i\eta)\rangle$, and $|X_{A}(\omega+ i\eta)\rangle$  
are included as target in the density-matrix renormalization
to build a new superblock at the next step.

\end{enumerate}
The robust finite-system DMRG algorithm must be used to perform
several sweeps through a lattice of fixed size.
Sweeps are repeated until the procedure has converged  to the
minimum of both functionals $E(\psi)$ and $W_{A,\eta}(\omega,\psi)$.

To obtain the dynamical correlation function 
$G_{A}(\omega+ i\eta)$ over a range of frequencies, one has to 
repeat this calculation for several frequencies $\omega$.
If the DDMRG calculations are performed independently, the computational
effort is roughly proportional to the number of frequencies.   
It is also possible to carry out a DDMRG calculation for
several frequencies simultaneously, including several states 
$|X_{A}(\omega+ i\eta)\rangle$ and $|Y_{A}(\omega+ i\eta)\rangle$ 
with different frequencies $\omega$ as target. 
The optimal number of different frequencies to be included
in a single calculation depends strongly on the problem studied and
the computer used.
As calculations for different frequencies are essentially 
independent, it would be easy and very efficient to perform
these calculations on a parallel computer.

If one performs a DDMRG calculation for two close frequencies
$\omega_1$ and $\omega_2$ simultaneously, it is possible to calculate
the dynamical correlation function for additional frequencies
$\omega$ between $\omega_1$ and $\omega_2$ without including
the corresponding states $|X_{A}(\omega+ i\eta)\rangle$ and 
$|Y_{A}(\omega+ i\eta)\rangle$ as target in the density-matrix
renormalization. 
This approach can significantly reduce the computer time necessary
to determine the spectrum over a frequency range but the results 
obtained for $\omega \neq \omega_1, \omega_2$ are less accurate and 
not always reliable, as the DMRG basis is optimized for the frequencies
$\omega_1$ and $\omega_2$ only.
(A similar technique is the calculation of spectra with the Lanczos 
algorithm in the DMRG basis optimized for a pair of correction vectors,
see Ref.~\onlinecite{till}.)
Alternatively, between the frequencies for which $G_A(\omega + i\eta)$
is determined directly with DDMRG, we can calculate the dynamical 
correlation function by interpolation using the DDMRG data for the 
function and its derivative.

If a complete spectrum $I_A(\omega + i\eta)$ has been obtained,
it is possible to calculate the moments of the spectral
distribution [the left-hand-side of Eq.~(\ref{sumrules})].
The first few moments [the right-hand-side of Eq.~(\ref{sumrules})] 
can be calculated accurately using the Lanczos DMRG 
method.~\cite{karen,till}
This provides an independent check of DDMRG results.     
[Note that only the first sum rule~(\ref{sumrules}) is satisfied 
exactly for $\eta > 0$.]

To calculate individual excited states contributing to the spectrum
in a given frequency range $(\omega_1,\omega_2$), 
one includes a minimization of $W_{A,\eta}(\omega,\psi)$
with respect to $\omega$ ($\omega_1 < \omega < \omega_2$)
in the third step of the DDMRG algorithm described above. 
In this case, $|Y_{A}(\omega_{\text{min}} + i\eta)\rangle$ and 
$|X_{A}(\omega_{\text{min}}+ i\eta)\rangle$ are included as target
in the sixth step.
The parameter $\eta$ must be much smaller than the distance
$E_{n+1} - E_{n}$ between two successive eigenstates contributing to the
dynamical correlation function or must decrease
during the calculation until the desired accuracy is obtained.
To make the procedure robust it is necessary to simultaneously target
a second correction vector $|\psi_{A}(\omega+ i\eta)\rangle$ 
with a fixed frequency and a parameter $\eta$ of the order of the
frequency range.
Typically, I use $\omega=(\omega_1+\omega_2)/2$ and 
$\eta=(\omega_2-\omega_1)/4$.

Because of the variational principle one naively expects that the DDMRG
results for $I_A(\omega +i\eta)$ must converge monotonically from below 
to the exact result as the number $m$ of density-matrix eigenstates is 
increased. 
In practice, the convergence is not always regular because of two 
approximations made to calculate the functional 
$W_{A,\eta}(\omega,\psi)$ in a DMRG basis.
First, the ground-state energy $E_0$ and the state $|A\rangle$ used 
in the definition~(\ref{functional}) of $W_{A,\eta}(\omega,\psi)$ 
are not known exactly but calculated with DMRG. 
If the number $m$ of density matrix eigenstates is increased, $E_0$ and
$|A\rangle$ are modified (they become progressively more accurate)
and the functional $W_{A,\eta}(\omega,\psi)$ is changed, which can 
affect its minimum arbitrarily.    
We also note that errors of the order of $\epsilon$ in $E_0$ or
$|A\rangle$ result in errors of the same order in $I_A(\omega +i\eta)$.
Therefore, to observe a regular convergence with increasing $m$ and to 
obtain accurate results for $I_A(\omega +i\eta)$, it is necessary in the
first place to determine the ground state and the state $|A\rangle$ 
with great precision (and thus to include the state $|A\rangle$ as 
target). 

To calculate the functional $W_{A,\eta}(\omega,\psi)$ in the third
step of the DDMRG algorithm, one needs an effective representation of 
the operator $(H-E_0-\omega)^2$ in the superblock subspace
\begin{equation}
[(H-E_0-\omega)^2]_{\text{eff}}  = O^{\dag} (H-E_0-\omega)^2 O \; ,
\end{equation}
where the operator $O$ represents the projection onto the
superblock subspace. 
For a typical many-body Hamiltonian $H$ such a calculation is 
excessively complicated and computationally intensive. 
Therefore, I calculate an effective representation of
$H$ only, $H_{\text{eff}} = O^{\dag} H O$, and assume that
\begin{equation}
[(H-E_0-\omega)^2]_{\text{eff}} \approx (H_{\text{eff}}-E_0-\omega)^2 
\label{substitution}
\end{equation}
to calculate $W_{A,\eta}(\omega,\psi)$ in the superblock subspace. 
This second approximation can cause a violation of the variational bound
$W_{A,\eta}(\omega,\psi) \geq -\pi \eta I_A(\omega +i\eta)$.
Fortunately, the substitution~(\ref{substitution}) has no significant
effect on the minimum of $W_{A,\eta}(\omega,\psi)$ if the state 
$(H-E_0-\omega) |Y_{A}(\omega + i\eta)\rangle \propto 
|X_{A}(\omega + i\eta)\rangle$ is accurately represented in the DMRG 
basis [i.e., if $O |X_{A}(\omega + i\eta)\rangle \approx 
|X_{A}(\omega + i\eta)\rangle$ for all superblock subspaces].
Therefore, to use the substitution~(\ref{substitution}) without loss 
of accuracy it is necessary and sufficient to include the state
$|X_{A}(\omega + i\eta)\rangle$ as a target in a DDMRG 
calculation, even if the real part of the dynamical correlation 
function is not calculated. 

In practice, for sufficiently large $m$ I have found that 
the absolute values of errors in a spectrum $I_A(\omega +i \eta)$ 
decrease systematically with increasing $m$. 
Therefore, it is possible to estimate the accuracy of a DDMRG 
calculation from the results obtained for different values of $m$ 
as one can do for static properties calculated with DMRG.
Moreover, DDMRG results for $I_A(\omega +i \eta)$ tend to be smaller 
than the exact result for almost all frequencies although they can
exceed it occasionally. 

Obviously, DDMRG is very similar to the correction vector
DMRG.~\cite{till,pati}
The same DMRG basis is built because the same target states are used 
in both methods.   
As numerical errors are usually dominated by the DMRG basis truncation,
the correction vector parts $|X_{A}(\omega + i\eta)\rangle$ 
and $|Y_{A}(\omega + i\eta)\rangle$ are calculated with the same 
precision $\epsilon$ in both methods for a given number $m$
of density-matrix eigenstates kept per block.  
Nevertheless, the variational formulation has two significant 
advantages.
First, the errors in the spectrum $I_A(\omega+i\eta)$, the excitation 
energies $E_n-E_0$, and the spectral weights are of the order
of $\epsilon^2$ instead of $\epsilon$ in the correction vector method,
as explained in Sec.~\ref{sec:principle}.
If one uses the Lanczos algorithm instead of Eq.~(\ref{dynamCF2})
in the correction vector DMRG, errors become even larger.
Therefore, DDMRG results are more accurate than those obtained with the
correction vector DMRG for a given number $m$ of density-matrix 
eigenstates. 
Second, a DDMRG calculation is essentially an application of the 
standard DMRG algorithm to the minimization of a different functional.
In particular, the numerical accuracy and computational effort 
are controlled by the sole parameter $m$ in an optimized DDMRG
calculation as in a ground-state DMRG calculation. 
The correction vector DMRG~\cite{till,pati} and Lanczos 
DMRG~\cite{karen,till} are significantly more complicated than the 
standard DMRG method. 
In particular, the numerical accuracy and computational effort depend 
significantly and sometimes unpredictably on the specific states 
(Lanczos vectors and correction vectors) included as target in the 
density-matrix renormalization.
Therefore, it is easier to implement and use DDMRG than the 
correction vector DMRG or the Lanczos DMRG.

\section{Finite-size scaling \label{sec:scaling}}

DDMRG allows us to calculate spectral functions of a correlated 
electron (or spin) system on a finite lattice with a broadening
given by the parameter $\eta > 0$. 
They have the generic form
\begin{equation}
I_{N, \eta}(\omega) =  \frac{1}{\pi} \sum_{n} A_n(N) 
\frac{\eta}{(\omega_n(N)-\omega)^2 + \eta^2} \; ,
\end{equation}
where $\omega_n(N)$ denotes the excitation energy and $A_n(N) > 0$ the 
spectral weight of the system eigenstates, and $N$ is the number of 
lattice sites.  
Such spectra are discrete for $\eta \rightarrow 0$ because there is 
only a finite number of eigenstates in a finite system. 
In the thermodynamic limit  a spectral function
\begin{equation}
I(\omega) =  \lim_{\eta \rightarrow 0} \lim_{N \rightarrow \infty}
I_{N, \eta}(\omega)   
\label{inflim}
\end{equation}
can contain discrete and continuous parts.
(It should be noted that the order of limits in the above
formula is important.)
To determine the properties of a dynamical spectrum $I(\omega)$ in the 
thermodynamic limit one has to analyze the scaling of the corresponding
spectra $I_{N, \eta}(\omega)$ as a function of system size.
Here I present a finite-size scaling technique for spectral functions 
calculated with a numerical method such as DDMRG. 

Computing both limits in Eq.~(\ref{inflim}) from numerical results
for $I_{N, \eta}(\omega)$ requires a lot of accurate data
for different values of $\eta$ and $N$ 
and can be the source of large extrapolation errors.
A much better approach is to use a broadening $\eta(N) >0$
which decreases with increasing $N$ and vanishes in the 
thermodynamic limit. 
The dynamical spectrum is then given by 
\begin{eqnarray}
I(\omega) & =  & \lim_{N \rightarrow \infty} I_{N, \eta(N)}(\omega)   
\label{scalinglimit} \\
& = & \lim_{N \rightarrow \infty} \frac{1}{\pi} \sum_{n} A_n(N) 
\frac{\eta(N)}{(\omega_n(N)-\omega)^2 + \eta^2(N)} \; . \nonumber  
\end{eqnarray}
From the existence of both limits in Eq.~(\ref{inflim})
it can be demonstrated that there exists a minimal broadening
$\eta_0(N) \geq 0$,
which decreases as a function of $N$ and converges to zero for
$N \rightarrow \infty$, such that the above equation is valid for all
functions $\eta(N)$ with $\eta(N) > \eta_0(N)$ and
$\lim_{N \rightarrow \infty} \eta(N) = 0$.
The function $\eta_0(N)$ depends on the frequency $\omega$ considered.
For a finite lattice with $N$ sites, let $M_{\omega, \epsilon}(N)$
be the number of excited states contributing to the spectral function
in a small interval of width $\epsilon$ around the 
frequency $\omega$ (i.e., $|\omega_n(N)-\omega| < \epsilon/2$).
If $M_{\omega, \epsilon}(N)$ remains finite for any $\epsilon > 0$ as 
$N \rightarrow \infty$, the spectral function $I(\omega)$ is discrete
at $\omega$ and $\eta_0(N)=0$. Equivalently, one can take the
$\eta \rightarrow 0$ limit first in Eq.~(\ref{inflim}).
If $M_{\omega, \epsilon}(N)$ diverges for all $\epsilon > 0$ as 
$N \rightarrow \infty$, the spectrum is dense at $\omega$ and 
a minimal broadening  $\eta_0(N) > 0$ is required for 
Eq.~(\ref{scalinglimit}) to be valid. 
For instance, $\eta_0(N)$ must be larger than the distance 
$\delta\omega = \omega_{n+1}(N)-\omega_n(N)$ between two consecutive 
excited states in the spectrum.
Note that while a continuous spectrum is obviously dense, a dense 
spectrum can be continuous or discrete. 
For instance, an infinite number of excited states with $A_n(N) > 0$ 
can converge to the same energy as $N \rightarrow \infty$.
This seems to happen for the optical conductivity associated
with an exciton in a open chain.~\cite{fabian} 

The function $\eta_0(N)$ depends naturally on the specific problem 
studied [i.e., the scaling of the energies $\omega_n(N)$ and spectral 
weights $A_n(N)$].  
For the optical conductivity of one-dimensional
correlated electron systems such as the Peierls-Hubbard model, 
I have found numerically that a sufficient
condition for all frequencies $\omega$ in a dense part of the optical
spectrum is
\begin{equation}
\eta  \geq \frac{c}{N} \; ,
\label{etacondition}
\end{equation}
where the constant $c$ is comparable to the width of the dynamical 
spectrum $I(\omega)$, which is finite in such lattice models.
Usually, one wants to keep the broadening $\eta$ as small as possible
because it reduces the resolution of the spectrum.
Therefore, I use
\begin{equation}
\eta(N)  = \frac{c}{N} 
\label{etascaling}
\end{equation}
in Eq.~(\ref{scalinglimit}) to analyze the finite-size scaling
of spectral functions calculated with DDMRG and to extrapolate the
finite-size results to the thermodynamic limit.
The condition~(\ref{etacondition}) has a very simple physical
interpretation. 
The spectral function $I_{N, \eta}(\omega)$ represents the dynamical 
response of the system over a time period $\sim 1/\eta$ after 
one has started to apply an external force.  
Typically, in a lattice model the spectral width is proportional to the
velocity of the excitations involved in the system response.
Thus the condition~(\ref{etacondition}) means that excitations are too 
slow to travel the full length $\sim N$ of the system in the time 
interval $\sim 1/\eta$ and do not ``sense'' that the system is finite.

An additional benefit of a broadening satisfying the 
condition~(\ref{etacondition}) 
is that the finite-system spectrum $I_{N,\eta}(\omega)$ becomes  
indistinguishable from the infinite-system spectrum with the same
broadening $\eta$ for relatively small $N$, 
\begin{equation}
I_{N,\eta}(\omega) \approx C_{\eta}[I(\omega)].
\label{comparison}
\end{equation}
Therefore, if one knows an exact or conjectured spectral function
$I(\omega)$ for an infinite system, its convolution with a 
Lorentzian of width $\eta$ can be compared directly with the numerical
results for the finite-system spectrum $I_{N,\eta}(\omega)$.
This approach has been applied successfully to the optical conductivity
of one-dimensional Mott insulators in Refs.~\onlinecite{eric} 
and \onlinecite{fabian}, 
and additional examples are presented in the next section. 

In practice, the extrapolation scheme~(\ref{scalinglimit})
works well at fixed frequency for the continuous parts and the 
non-dense discrete parts of a spectrum $I(\omega)$ only.
To detect singularities in $I(\omega)$ and determine their properties, 
it is generally easier to analyze the scaling of maxima in 
$I_{N,\eta}(\omega)$ or in its derivative as $N \rightarrow \infty$
and $\eta \rightarrow 0$.
To perform this scaling analysis one can use a size-dependent 
broadening $\eta(N)$ such that Eq.~(\ref{scalinglimit}) is valid 
and $I_{N,\eta}(\omega)$ is a good approximation of 
$C_{\eta}[I(\omega)]$ around the maximum.
Then the scaling of a maximum in $I_{N,\eta}(\omega)$ for $\eta(N) 
\rightarrow 0$ gives the scaling of the corresponding maximum in 
$C_{\eta}[I(\omega)]$ for $\eta \rightarrow 0$.
Here I discuss some examples of this technique which are useful
for the analysis of the Peierls-Hubbard model optical conductivity
presented in the next section.
[One should also note that to detect the presence of a gap between
two bands in an infinite-system spectrum  $I(\omega)$,
it is often faster and more reliable to investigate the scaling of 
minima in $I_{N,\eta}(\omega)$ for $\eta(N) \rightarrow 0$
than to perform extrapolations at fixed frequencies.]

First, we consider an infinite-system spectrum with a peak in a
continuous band,
\begin{equation}
I(\omega) = I_0 \delta(\omega - \omega_0) + I_{\text{cont}}(\omega)
\end{equation}
for $|\omega - \omega_0| < \Lambda$, where $I_0 > 0$ and 
$I_{\text{cont}}(\omega)$ is a continuous function.
It is easy to show that for $\eta \ll \Lambda$ the maximum of 
$C_{\eta}[I(\omega)]$ diverges as $I_0/(\pi \eta)$ and that the position
of the maximum converges to $\omega_0$ for $\eta \rightarrow 0$.
The maximum of the corresponding finite-system spectra 
$I_{N,\eta}(\omega)$ has the same scaling 
properties for $\eta(N) \rightarrow 0$.
Therefore, it is possible to detect such a $\delta$-peak and determine
its weight $I_0$ in an infinite-system spectrum, even if $I_0$ is only 
a small fraction of the total spectral weight of $I_{\text{cont}}$. 

Second, we consider an infinite-system spectrum with a power-law
divergence at the band edge
\begin{equation}
I(\omega) = I_0 \theta(\omega - \omega_0) |\omega - \omega_0|^{-\alpha} 
\end{equation}
for $|\omega - \omega_0| < \Lambda$, where $\theta(x) = 0$ for $x < 0$
and $\theta(x) =1$ for $x > 0$, $I_0 > 0$, and $0 < \alpha < 1$.
One can show that for $\eta \ll \Lambda$ the maximum of
$C_{\eta}[I(\omega)]$ diverges as $\eta^{-\alpha}$ and 
that the position of the maximum converges to $\omega_0$ from above for
$\eta \rightarrow 0$.
Again, the maximum of the corresponding finite-system spectra 
$I_{N,\eta}(\omega)$ has the same scaling
properties for $\eta(N) \rightarrow 0$.
Thus it is possible to detect such a singularity and determine the
exponent $\alpha$ from the finite-system numerical data.

Third, we consider a continuous infinite-system spectrum
with a singularity in its derivative
\begin{equation}
I(\omega) = I_0 \theta(\omega - \omega_0) |\omega - \omega_0|^{\alpha} 
\end{equation}
for $|\omega - \omega_0| < \Lambda$, where the function $\theta(x)$
and the constants $I_0$ and $\alpha$ are as in the previous example.
Here the derivative of $C_{\eta}[I(\omega)]$ has a maximum which 
diverges as $\eta^{\alpha-1}$ for $\eta \ll \Lambda$ while its position
converges to $\omega_0$ from above as $\eta \rightarrow 0$.  
The maximum in the derivative of $I_{N,\eta}(\omega)$ has the same 
scaling properties for $\eta(N) \rightarrow 0$.
Therefore, it is possible to determine the exponent $\alpha$ from
the finite-system numerical data in this case too.

Finally, we consider a special function representing a continuous 
spectrum above a gap and a truncated divergence close to the band edge
\begin{equation}
I(\omega) =  I_0 \theta(\omega -\omega_0) 
\frac{2\sqrt{\omega_0 |\omega -\omega_0|}}
{\gamma \omega_0 +|\omega -\omega_0|}
\label{typicalOC}
\end{equation}
for $|\omega - \omega_0| < \Lambda$, where $\gamma$ is a constant such 
that $0 \leq \gamma<\Lambda/\omega_0$.
The function $\theta(x)$ and the other constants are as in the previous
examples.  
For $\gamma >0$ this spectrum vanishes as a square-root at the band edge
$\omega_0$, goes trough a maximum $I_0/\sqrt{\gamma}$ at 
$\omega = (1+\gamma) \omega_0$ then decreases monotonically.  
For $\gamma \ll 1$ the maximum is very sharp and close to the band edge,
and $I(\omega)$ appears to diverge as $1/\sqrt{|\omega -\omega_0|}$
at higher energy $ \omega > (1+\gamma) \omega_0$.
The continuous vanishing of $I(\omega)$ at the band edge is
apparent only in a small frequency range $\omega_0 < \omega < (1+\gamma)
\omega_0$. 
As $\gamma \rightarrow 0$ this maximum becomes a square-root singularity
at $\omega_0$. 
A qualitatively similar behavior is often found in the optical 
conductivity of one-dimensional insulators (see the discussion in the 
next section).  
Obviously, the maximum of $C_{\eta}[I(\omega)]$ tends to 
$I_0/\sqrt{\gamma}$ and its position converges to 
$\omega = (1+\gamma) \omega_0$ for $\eta \rightarrow 0$.
For $\gamma < 1$, however, the convergence of the maximum becomes 
apparent only for $\eta \ll \gamma \omega_0$. 
For larger $\eta$ the maximum appears to diverge as $1/\sqrt{\eta}$.
Similarly, the maximum of the derivative diverges as $1/\sqrt{\eta}$
for $\eta \rightarrow 0$ as discussed the previous example.
In the present case, however, this scaling is not observed as soon as 
$\eta \ll \Lambda$ but only if $\eta \ll \gamma \omega_0$.
The finite-system spectrum $I_{N,\eta}(\omega)$ and its derivative have 
the same scaling properties for $\eta(N) \rightarrow 0$.
Therefore, with the scaling analysis of finite-system spectra 
it is possible to distinguish an infinite-system spectrum with a 
truncated divergence above the band edge from a spectrum with a 
real divergence at the band edge, provided that one can do calculations
with a resolution $\eta(N) \ll \gamma \omega_0$.

In summary, the dynamical spectrum of an infinite system can be
determined accurately and efficiently from numerical data for 
finite-system spectra using a size-dependent broadening $\eta(N)$.
The broadening $\eta(N)$ must be larger than a minimal broadening 
$\eta_0(N)$, which depends on the system investigated and can vary
with the frequency.
Often this broadening conceals the finite-size effects and one
can directly compare finite-system spectra
to analytical results for infinite systems using a convolution with
a Lorentzian distribution, see Eq.~(\ref{comparison}).
If this comparison is not possible or not sufficient, specific
points of the spectrum can be extrapolated to the thermodynamic limit
using Eq.~(\ref{scalinglimit}).
Finally, the scaling of maxima in finite-system spectra or their
derivatives [for $\eta(N) \rightarrow 0$] allows us to find and analyze 
singularities in the infinite-system spectrum. 
For one-dimensional correlated electron systems a sufficient
condition for the minimal broadening is given by 
Eq.~(\ref{etacondition}) and one can use a size-dependent 
broadening~(\ref{etascaling}).

\section{Optical conductivity of the Peierls-Hubbard model 
\label{sec:model}}

In this section I apply the DDMRG method and the finite-size
scaling analysis to the optical conductivity
of the one-dimensional Peierls-Hubbard model.~\cite{dionys}
This model is defined by the Hamiltonian
\begin{equation}
H = T + U \sum_{l=1}^N \left(n_{l,\uparrow}-\frac{1}{2}\right)
\left(n_{l,\downarrow}-\frac{1}{2}\right)  
\label{hamiltonian}
\end{equation}
with 
\begin{equation}
T = - \sum_{l;\sigma} \left(t-(-1)^l \frac{\Delta}{2} \right )
\left(c_{l,\sigma}^+c_{l+1,\sigma} + c_{l+1,\sigma}^+c_{l,\sigma}\right)
\; .
\label{kinetic}
\end{equation}
It describes electrons with spin $\sigma=\uparrow,\downarrow$ 
which can hop between neighboring sites in a lattice with an even number
$N$ of sites.
In Eq.~(\ref{kinetic}) the index $l$ runs from 1 to $N-1$ for an open 
chain and from 1 to $N$ if periodic boundary conditions are used. 
Here $c^+_{l,\sigma}$, $c_{l,\sigma}$ are creation and annihilation 
operators for electrons with spin $\sigma$ at site $l$ and 
$n_{l,\sigma}= c^+_{l,\sigma}c_{l,\sigma}$ are the corresponding 
density operators. 
The hopping integral $t > 0$ gives rise to a single-electron band of 
width $4t$.
The dimerization parameter $0 \leq |\Delta| \leq 2t$ determines
the strength of the periodic lattice potential generated by the Peierls
instability.
(For a finite open chain I only use $\Delta \geq 0$ to avoid spurious
excitations at the chain ends.)
The Coulomb repulsion is mimicked by a local Hubbard interaction 
$U\geq 0$.
The number of electrons equals the number of lattice sites. 

The ground state, single-particle charge gap and spin gap of this
system can be calculated with great accuracy on lattices with up to 
$N \sim 10^3$ sites using DMRG. 
The single-particle charge gap is given by  
\begin{equation}
E_c(N) =  E_0(N+1) + E_0(N-1) - 2 E_0(N) \; ,
\label{chargegap}
\end{equation}
where $E_0(M)$ denotes the ground-state energy for $M$ electrons
in a $N$-site system.
For an even number $N$ of sites and electrons the Peierls-Hubbard model
ground state is a singlet~\cite{lieb} and the spin gap is given by
\begin{equation}
E_s(N) =  E_0(S_z=\hbar) - E_0(S_z=0) \; ,
\end{equation}
where $S_z$ is the the $z$-component of the total spin and 
$E_0(S_z)$ is the ground-state energy for a fixed value of $S_z$.

Spectroscopy with electromagnetic radiation is a common experimental 
probe of solid-state materials.
The linear optical absorption is proportional 
to the real part $\sigma_1(\omega)$ of the optical conductivity.
For $\omega > 0$, $\sigma_1(\omega)$
is related to the imaginary part of the current-current 
correlation function $G_{J} (\hbar \omega+ i \eta)$ by
\begin{eqnarray}
\sigma_1(\omega) & = & \frac{\pi}{N a \omega} \lim_{\eta \rightarrow 0}
\text{Im} \ G_{J} (\hbar \omega+ i \eta)  \label{sigma1} \\
& = &  \frac{\pi}{Na \omega} \sum_n 
|\langle \psi_0 |J| \psi_n\rangle|^2 \ \delta(\hbar\omega+E_0- E_n) \; .
\nonumber
\end{eqnarray}
Here $|\psi_0\rangle$ is the ground state of the
Hamiltonian $H$, $|\psi_n\rangle (n > 1)$ are the other eigenstates
of $H$, and $E_0$, $E_n$ are their respective eigenenergies. 
In this model the current operator is 
\begin{equation}
J=\frac{iae}{\hbar}  
\sum_{l;\sigma} \left(t-(-1)^l \frac{\Delta}{2} \right )
\left( c_{l,\sigma}^+c_{l+1,\sigma} 
- c_{l+1,\sigma}^+c_{l,\sigma} \right) \; ,
\end{equation} 
where $a$ is the lattice constant, $-e$ is the charge of an electron, 
and the index $l$ takes the same values depending on the boundary
conditions as in Eq.~(\ref{kinetic}). 
Note that this is the natural definition of the current operator for 
both types of boundary conditions. 
The Parzen filter used for open boundary conditions in other 
works~\cite{till2,brune} is not necessary and thus not used in this 
work.

In an open chain the optical absorption is also related to the
dynamical polarizability $\alpha(\omega)$, which is given by the 
imaginary part of the dipole-dipole correlation function
$G_{D} (\hbar \omega+ i \eta)$, 
\begin{eqnarray}
\alpha(\omega) & = & \frac{\pi}{N a} \lim_{\eta \rightarrow 0}
\text{Im} \ G_{D} (\hbar \omega+ i \eta)  
\label{alpha} \\
& = &  \frac{\pi}{Na} \sum_n |\langle \psi_0 |D| \psi_n\rangle|^2 \
\delta(\hbar\omega+E_0- E_n) \; .
\nonumber     
\end{eqnarray}
For a constant lattice spacing $a$ the dipole operator is
\begin{equation}
D=-ea 
\sum_{l=1}^N l \left(n_{l} - 1 \right )  \;   
\end{equation} 
with $n_{l} = n_{l, \uparrow} + n_{l, \downarrow}$.
Using the relation $J = -\frac{i}{\hbar} [D,H]$ one easily proves that
\begin{equation}
\sigma_1(\omega) = \omega \alpha(\omega)  
\label{sigmaalpha}
\end{equation}
and  
\begin{equation}
\sigma_1(\omega) =  \frac{\pi}{N a \hbar} \lim_{\eta \rightarrow 0}
\text{Im} \{ (\hbar \omega+ i\eta) \ G_{D} (\hbar \omega+ i \eta)\} \; .
\label{sigma1bis}
\end{equation}

The optical conductivity can be calculated with the DDMRG method
described in this paper.
For an open chain Eqs.~(\ref{sigma1}),(\ref{alpha}), 
and (\ref{sigma1bis}) provide us with three different approaches.
First, one can calculate the imaginary part of the current-current
correlation function with DDMRG and use Eq.~(\ref{sigma1}) to obtain 
the convolution of the reduced optical conductivity
\begin{eqnarray}
C_{\eta}[\omega \sigma_1(\omega)] & = & 
\frac{\pi}{N a} \text{Im} \ G_{J} (\hbar \omega+ i \eta)  
\label{Crsigma} \\
& = &  \frac{1}{Na} \sum_n 
\frac{\eta |\langle \psi_0 |J| \psi_n\rangle|^2} 
{(\hbar\omega+E_0- E_n)^2+\eta^2} \; .
\nonumber             
\end{eqnarray}
This is also the only approach possible with periodic boundary 
conditions.
Second, one can calculate the imaginary part of the dipole-dipole 
correlation function with DDMRG and use Eq.~(\ref{alpha}) 
to obtain the convolution of the dynamical polarizability
\begin{eqnarray}
 C_{\eta}[\alpha(\omega)] & = & \frac{\pi}{N a} 
\text{Im} \ G_{D} (\hbar \omega+ i \eta)  
\label{Calpha} \\
& = &  \frac{1}{Na} \sum_n 
\frac{\eta |\langle \psi_0 |D| \psi_n\rangle|^2}
{(\hbar\omega+E_0- E_n)^2+\eta^2} \; .
\nonumber 
\end{eqnarray}
The optical conductivity is then given by the 
relation~(\ref{sigmaalpha}).
Finally, one can calculate the complete dipole-dipole correlation 
function with DDMRG and use Eq.~(\ref{sigma1bis}) to obtain
the convolution of $\sigma_1(\omega)$ directly
\begin{eqnarray}
C_{\eta}[\sigma_1(\omega)] & = & \frac{\pi}{N a \hbar} 
\text{Im} \{ (\hbar \omega+ i\eta) \ G_{D} (\hbar \omega+ i \eta)\}
\label{Csigma1} \\
& = & \frac{1}{Na\hbar} \sum_n 
\frac{\eta |\langle \psi_0 |D| \psi_n\rangle|^2 (E_n -E_0)} 
{(\hbar\omega+E_0- E_n)^2+\eta^2} \; . \nonumber   
\end{eqnarray}
$C_{\eta}[\sigma_1(\omega)]$ can also be formulated in terms of
the current matrix elements $|\langle \psi_0 |J| \psi_n\rangle|^2$
[see Eq.~(4) of Ref.~\onlinecite{fabian}].
Although the real part of a dynamical correlation function is used in 
Eq.~(\ref{Csigma1}) to calculate the optical conductivity,
its relative contribution to $C_{\eta}[\sigma_1(\omega)]$ is of the 
order $(\eta/t)^2$.
Therefore, the numerical precision is not significantly reduced by the 
lower accuracy of DDMRG for the real part of dynamical correlation 
functions. 

Clearly, all three approaches give the same spectrum $\sigma_1(\omega)$
for $\eta \rightarrow 0$. 
In DDMRG calculations with $\eta > 0$, however, 
they are not equivalent.
First, I have found that it is easier to calculate the dipole-dipole
correlation function $G_{D} (\hbar \omega+ i \eta)$ than
the current-current correlation function $G_{J} (\hbar \omega+ i \eta)$,
except for very strong coupling $U \gg t$.
Second, the finite-size scaling [using a size-dependent 
broadening~(\ref{etascaling})] is different for the three optical 
spectra $C_{\eta}[\omega \sigma_1(\omega)]/\omega$, 
$\omega C_{\eta}[\alpha(\omega)]$, and $C_{\eta}[\sigma_1(\omega)]$,
especially for very small and very large frequencies $\omega$.
Usually, $C_{\eta}[\sigma_1(\omega)]$ is the best approximation
to $\sigma_1(\omega)$ but at low energy ($\hbar \omega < t$)
it can be more convenient to use $\omega C_{\eta}[\alpha(\omega)]$ 
while at high energy ($\hbar \omega \gg t$)
I prefer $C_{\eta}[\omega \sigma_1(\omega)]$.

To calculate the optical spectrum of the Peierls-Hubbard model I have 
used the third approach, Eq.~(\ref{Csigma1}), in most cases. 
Thus $C_{\eta}[\sigma_1(\omega)]$ is shown in the figures of this paper
unless I state explicitly otherwise.
Only optical spectra calculated with open boundary conditions are 
presented. 
As the size-dependent broadening~(\ref{etascaling}) conceals most of the
finite-size effects, spectra calculated with periodic boundary 
conditions would be almost identical. 
In all figures showing optical spectra I set $a=e=\hbar=t=1$.
Thus $\sigma_1(\omega)$ is shown in units of $e^2a/\hbar$, 
$\omega \sigma_1(\omega)$ in units of $e^2a t/\hbar^2$, and
the frequency $\omega$ in units of $\hbar/t$.

The sum rules~(\ref{sumrules}) take a simple form for the optical 
conductivity in the Peierls-Hubbard model with open boundary conditions
\begin{subequations}
\label{OCsums}
\begin{eqnarray}
\frac{\hbar}{\pi} \int_0^\infty d\omega \omega \sigma_1(\omega)&=&
\frac{1}{Na} \ \langle \psi_0|J^2|\psi_0\rangle ,
\label{sumJ}  \\
\frac{\hbar}{\pi} \int_0^\infty d\omega \sigma_1(\omega) \; \; &=&
\frac{-a e^2}{2N\hbar} \ \langle \psi_0|T|\psi_0\rangle ,
\label{sumT}\\
\frac{\hbar}{\pi} \int_0^\infty d\omega \frac{\sigma_1(\omega)}{\omega}
\; \, &=& \frac{1}{Na} \, \langle \psi_0| D^2 |\psi_0\rangle\ .
\label{sumD}
\end{eqnarray}
\end{subequations}
To prove the second sum rule~(\ref{sumT}) one uses the relation
$[D,J] = -i a^2 e^2 T/\hbar $.  
With periodic boundary conditions, only the first two sum rules 
remain valid. In the second sum rule, however,
one must take into account the coherent part of the conductivity
at $\omega =0$ and the proof is more complicated
than for an open chain.

The right-hand-side of Eq.~(\ref{OCsums}) can be calculated accurately
with the ground-state or Lanczos DMRG method.
Note that for $\eta > 0$ an optical spectrum calculated from 
Eq.~(\ref{Crsigma}), (\ref{Csigma1}), or (\ref{Calpha}) 
exactly fulfills the sum rule (\ref{sumJ}), (\ref{sumT}), or
(\ref{sumD}), respectively. 
For the DDMRG results presented in this paper, the sum rules are 
fulfilled within a few percents.

The optical gap $E_{\text{opt}}(N)$ is defined as the excitation 
energy ($E_n - E_0$) of the lowest eigenstate contributing to the 
optical conductivity (i.e., $\langle \psi_n|J| \psi_0\rangle \neq 0$)
in a $N$-site system.
$E_{\text{opt}}(N)$ can be calculated with the DDMRG method for
individual excited states described in Sec.~\ref{sec:ddmrg}.
As the Peierls-Hubbard Hamiltonian~(\ref{hamiltonian}) has a
particle-hole symmetry the optical gap can also be determined
using the symmetrized DMRG method.~\cite{ramasesha} 
As expected, both approaches give the same results for 
$E_{\text{opt}}(N)$ within numerical errors.
In the thermodynamic limit ($N \rightarrow \infty$) I have found that 
the optical gap $E_{\text{opt}}$ is equal to the single-particle charge
gap $E_c$ (\ref{chargegap}) for all $U \geq 0$ and $2t > \Delta \geq 0$.
[In the dimer limit ($\Delta=2t$) the Hamiltonian~(\ref{hamiltonian}) 
describes independent dimers: the optical weight is concentrated
in a single peak corresponding to Frenkel excitons localized on 
a dimer and the other (delocalized) excitations above and below this 
peak carry not optical weight, thus $E_c < E_{\text{opt}}$.]
In a finite system or with additional electronic 
interactions~\cite{fabian} the single-particle charge gap $E_c$ can be 
different from the optical gap $E_{\text{opt}}$.

All DMRG methods have a truncation error which is 
reduced by increasing the number~$m$ of retained density matrix 
eigenstates (for more details, see Refs.~\onlinecite{steve} 
and~\onlinecite{dmrgbook}).  
Varying $m$ allows one to compute
physical quantities for different truncation errors and thus to obtain
error estimates on these quantities.
For some quantities, especially eigenenergies, it is possible 
to extrapolate the results to the limit of vanishing
truncation error and thus to achieve a greater accuracy.
I have systematically used these procedures to
estimate the precision of my numerical calculations
and adjusted the maximal number $m$ of density matrix 
eigenstates to reach a desired accuracy.
This is especially important for DDMRG calculations as truncation 
errors in dynamical spectra can greatly vary as a function of the 
frequency $\omega$ for fixed $m$. 
In this work the largest number of density matrix eigenstates 
used is $m=600$.
For all numerical results presented here DMRG truncation errors are 
negligible.

In the following three subsections I demonstrate the finite-size 
scaling technique and the accuracy of DDMRG on three special limits
of the Peierls-Hubbard model.
Then the optical conductivity of a Mott-Peierls insulator
is presented and discussed in the last subsection.

\subsection{Peierls insulator \label{sec:peierls}}

For $U=0$ the Hamiltonian~(\ref{hamiltonian}) describes a system
of independent electrons, which can be solved exactly for any
value of $\Delta$, boundary conditions, or lattice size.
This provides us with a perfect test case for the DDMRG method.
I have checked that DDMRG can reproduce the optical spectrum
of this system on lattices with several hundred sites,
for any frequency $\omega$, and with relative errors as small
as $10^{-4}$ using only a few hundred density-matrix eigenstates.
This demonstrates that one can obtain almost exact results for the 
optical conductivity of finite one-dimensional systems such as the 
Peierls-Hubbard model using DDMRG.

\begin{figure}
\includegraphics[width=8cm]{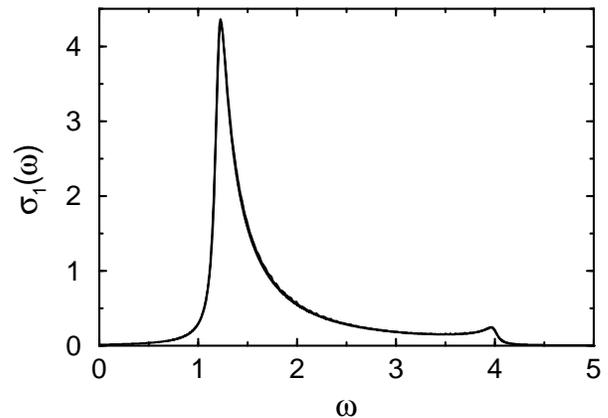}
\caption{ \label{fig1}
Optical conductivity of a Peierls insulator with $\Delta=0.6t$ and 
a broadening $\eta/t=0.05$.  
Both the DDMRG result for a 128-site chain and the exact 
result~(\ref{peierlsOC}) in the thermodynamic limit are shown.  
}
\end{figure}

In the thermodynamic limit the Hamiltonian~(\ref{hamiltonian}) 
describes a Peierls (band) insulating phase for  
$\Delta \neq 0$ and $U=0$.
The optical gap $E_{\text{opt}}$, the charge gap $E_c$, and the spin
gap $E_s$ equal $2|\Delta|$.
Optical excitations are made of one hole in the valence band and
one electron in the conduction band.
The optical conductivity is given by
\begin{equation}
\sigma_1(\omega) =  \frac{a e^2 (2\Delta)^2(4t)^2}
{2 \hbar(\hbar \omega)^2 \sqrt{[(\hbar \omega)^2-(2\Delta)^2]
[(4t)^2-(\hbar \omega)^2]}}
\label{peierlsOC}
\end{equation}
for $2|\Delta| < \hbar \omega <4t$ and is zero elsewhere.~\cite{florian}
This optical spectrum contains a single band of width
$4t-2|\Delta|$ with  square-root divergences at both band edges.
These divergences are a typical feature of a one-dimensional band 
insulator.
The convolution of Eq.~(\ref{peierlsOC}) with a Lorentzian distribution
of width $\eta/t=0.05$ is shown in Fig.~\ref{fig1} for $\Delta=0.6t$.
Both divergences are replaced by maxima at $\hbar \omega 
\approx 2 \Delta = 1.2t$ and $\hbar \omega \approx 4t$.
In Fig.~\ref{fig1} I also show the optical conductivity calculated with
DDMRG on a 128-site lattice with the same broadening.
We see that the finite-system optical spectrum is 
indistinguishable from the infinite-system spectrum.
The broadening $\eta/t=0.05$ satisfies the 
condition~(\ref{etacondition}) and thus conceals the finite-size 
effects as discussed in Sec~\ref{sec:scaling}.
In this case a broadening $\eta(N)/t=6.4/N$ is enough because the 
spectrum band width is smaller than $4t$.

With this size-dependent broadening $\eta(N)$ one can use 
Eq.~(\ref{scalinglimit}) 
to extrapolate the finite-size DDMRG results to the thermodynamic limit.
For instance, for $\hbar\omega=2.6t$ I have obtained 
$\sigma_1(\omega)= 0.245$ (in units of $ae^2/\hbar$) using data for 
systems with up to $N=256$ sites [i.e., with a broadening down to 
$\eta(N)/t = 0.025$], in excellent agreement with the exact result
0.243.
If we did not know the exact result~(\ref{peierlsOC}),
we could nevertheless determine the existence of square-root 
divergences at both band edges using a scaling analysis of the maxima 
in the DDMRG spectra. 
For instance, the height of the low-energy maximum (close to
$\hbar \omega = 1.2t$)  diverges as $1/\sqrt{\eta}$ for
$\eta(N) \rightarrow 0$ [see Fig.~\ref{fig2}(a)].
Moreover, the position of the maximum tends from above to the optical 
gap
$E_{\text{opt}} = 2 \Delta = 1.2 t$ for $N \rightarrow \infty$
[see Fig.~\ref{fig2}(b)].
As explained in Sec.~\ref{sec:scaling} these scaling properties 
correspond to a square-root divergence at the band edge.
Figure~\ref{fig2}(b) also shows the finite-size optical gaps 
$E_{\text{opt}}(N)$ calculated with the DDMRG method for individual 
excited states. 
They tend to the exact result $E_{\text{opt}} = 1.2 t$ for $N 
\rightarrow \infty$ as expected.

\subsection{Mott-Hubbard insulator \label{sec:hubbard}}

\begin{figure}
\includegraphics[width=7cm]{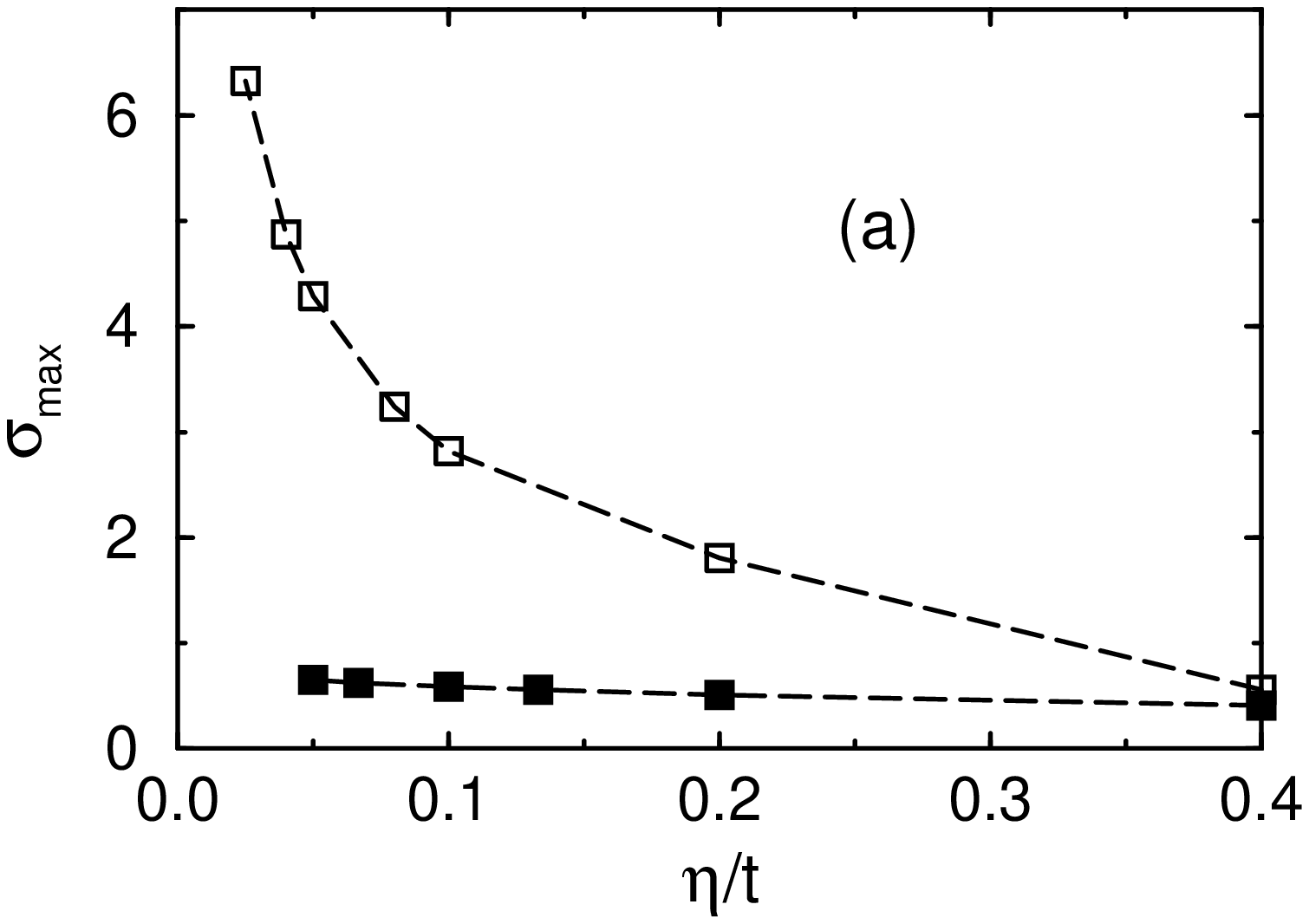}
\includegraphics[width=7cm]{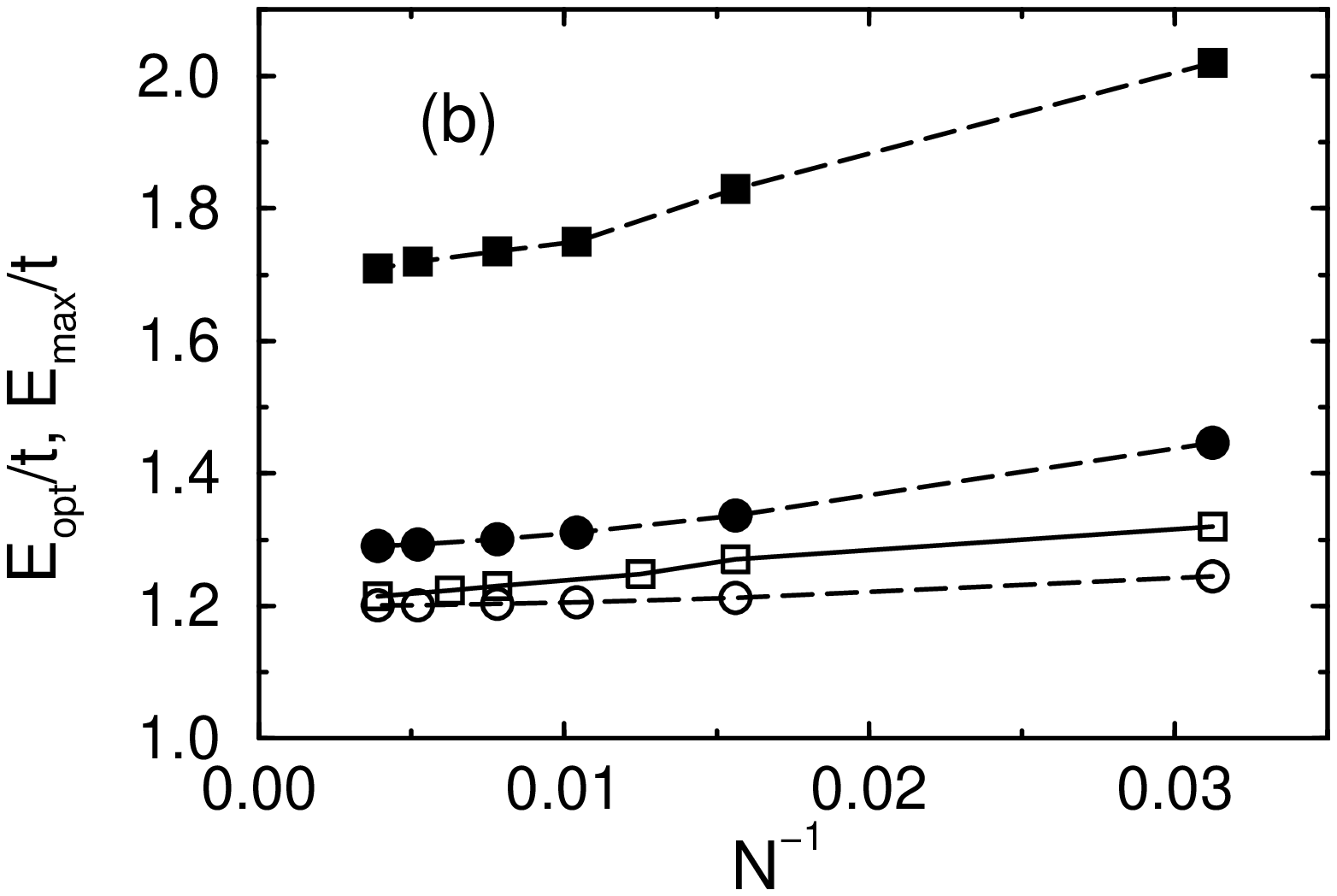}
\caption{ \label{fig2}
(a) Maximum $\sigma_{\text{max}}$ of the optical spectrum 
$\sigma_1(\omega)$ as a function of the broadening $\eta(N)$. 
(b) Position of the spectrum maximum $E_{\text{max}} = \hbar 
\omega_{\text{max}}$ (square) and
optical gap $E_{\text{opt}}$ (circle) as a function of system size $N$. 
In both figures filled symbols correspond to the Mott-Hubbard insulator
[$U=4t, \eta(N)N=12.8t$] and open symbols to the Peierls insulator 
[$\Delta=0.6t, \eta(N)N= 6.4t$].
}
\end{figure}

For $\Delta = 0$ the Peierls-Hubbard model~(\ref{hamiltonian})
becomes the one-dimensional Hubbard model at half filling.
For $U > 0$ this model describes a Mott-Hubbard insulator
with gapless spin excitations.~\cite{lieb2}
The optical conductivity of this system has recently been determined 
using DDMRG and analytical methods.~\cite{eric}
Here I only summarize the most important results and give more 
information about the finite-size scaling analysis carried out 
in this previous work. 

In the half-filled Hubbard model an optical excitation is made of a pair
of spinless bosonic excitations carrying opposite charges in the lower 
(holon) and upper (doublon or anti-holon) Hubbard bands, respectively.
As in a Peierls insulator, the optical spectrum consists of a single 
band but its width is larger, about $8t$.
A second distinctive feature of this spectrum is a square-root vanishing
$\sigma_1(\omega) \sim \sqrt{\hbar \omega - E_{\text{opt}}}$ at the 
band threshold $E_{\text{opt}}$.
There is also a tiny peak in the middle of the band,
at least for $U \geq 4t$.

\begin{figure}
\includegraphics[width=8cm]{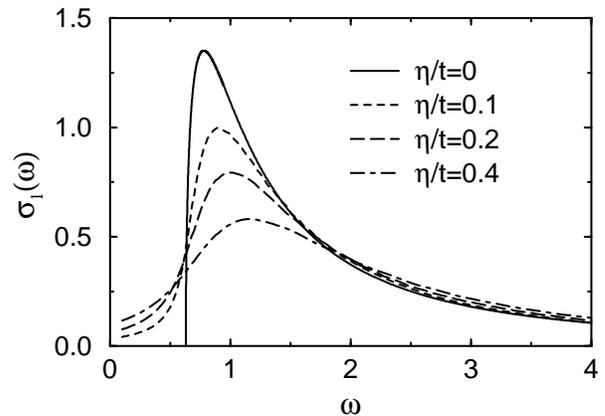}
\caption{ \label{fig3}
Optical conductivity of the Hubbard model with $U=3t$ for
several values of $\eta$. 
Results for $\eta(N) > 0$ have been calculated with DDMRG on $N$-site
lattices with $\eta(N) N =12.8t$.
For $\eta=0$ the field-theoretical result for an infinite system is 
shown.
}
\end{figure}

In Ref.~\onlinecite{eric} it is shown that for weak coupling
($U \leq 3t$) and in the strong-coupling limit 
($U/t \rightarrow \infty$) the finite-system optical spectra
calculated with DDMRG agree perfectly with the analytical results
obtained in the thermodynamic limit using a field-theoretical approach
and a strong-coupling analysis, respectively.  
For instance, Fig.~\ref{fig3} shows the low-energy parts of
DDMRG spectra calculated for three different lattice sizes at $U=3t$ 
and the corresponding field-theoretical spectrum for an infinite system.
A size-dependent broadening $\eta(N)/t=12.8/N$ is used in this case.
One clearly sees the convergence of the finite-size spectra toward
the field-theoretical result as $\eta(N)$ decreases.
To make a quantitative comparison one can calculate the convolution of
the field-theoretical spectrum with a Lorentzian of width  $\eta$
satisfying the condition~(\ref{etacondition}) as discussed in 
Sec.~\ref{sec:scaling}.
One finds then that finite-size effects are completely concealed by
the broadening even for relatively small system sizes.
For instance, it is shown in Fig.~3 of Ref.~\onlinecite{eric} that the 
low-energy optical spectrum calculated on a 128-site lattice for 
$U=3t$ is indistinguishable from the field-theoretical spectrum with 
the same broadening $\eta/t=0.1$.
In the strong-coupling limit DDMRG and analytical results agree even 
better and finite-size effects are no longer visible for systems
as small as $N=32$.  

For other coupling strengths ($4 \leq U/t  < \infty$) it is necessary
to analyze the scaling of the finite-system DDMRG spectra to determine
the optical conductivity of the Hubbard model in the thermodynamic
limit.
Using numerical results for lattices with up to $N=256$ sites
[i.e., a resolution $\eta(N)/t$  down to 0.05], I have found
that for all $U/t$ the optical conductivity at the lower band edge
has the qualitative behavior described by Eq.~(\ref{typicalOC}): 
$\sigma_1(\omega)$ vanishes as $\sqrt{\hbar \omega - E_{\text{opt}}}$
at the band threshold and there is a maximum in $\sigma_1(\omega)$
at a frequency $\omega = (1 + \gamma) E_{\text{opt}}/\hbar$, where
$\gamma$ is a small number.
Field theory  predicts the same behavior with $\gamma \approx 0.24$
in the weak-coupling limit while the strong-coupling analysis 
gives a maximum at $\hbar \omega = U = E_{\text{opt}} + 4t$, and thus
$\gamma$ vanishes as $t/U$ for $U \gg t$.
The distance $\gamma E_{\text{opt}}$ between the spectrum threshold and
the maximum increases with $U/t$.
For $U \geq 4t$ this distance is large enough to determine
the finite-size scaling of the lower band edge using systems 
with up to $N=256$ sites. 
As an example, Fig.~\ref{fig2}(a) shows the low-energy maximum in
the optical spectrum $\sigma_1(\omega)$ calculated with DDMRG
for $U=4t$ as a function of $\eta(N)$.
The contrast between the Mott-Hubbard insulator and the Peierls 
insulator is striking and the maximum in the Mott-Hubbard insulator 
optical spectrum clearly tends to a constant for 
$\eta(N) \rightarrow 0$.
[For $U=4t$ the optical gap of the Hubbard model 
is comparable to that of the Peierls insulator 
with $\Delta=0.6t$, so that a direct comparison of both systems is 
relevant.]
In Fig.~\ref{fig2}(b) one sees that the finite-size optical gaps 
calculated with DDMRG converge to the exact result~\cite{lieb2} 
$E_{\text{opt}} =  1.287t$ in the thermodynamic limit,
but the maximum tends to a higher energy 
$\hbar \omega \approx 1.7t$.
Therefore, one can conclude that there is no divergence
at the optical conductivity threshold $\hbar \omega = E_{\text{opt}}$.
Moreover, it is possible to confirm that $\sigma_1(\omega)$ vanishes
as $\sqrt{\hbar \omega - E_{\text{opt}}}$ at the lower band edge using
either a similar scaling analysis for the derivative of DDMRG spectra
or a direct comparison with the convolution of functions such as 
Eq.~(\ref{typicalOC}) or the field-theoretical optical 
spectrum.~\cite{eric} 

For very weak coupling one would need to calculate
$\sigma_1(\omega)$ for very large system sizes in order to perform 
the same scaling analysis.
Because $E_{\text{opt}}$ vanishes exponentially with $U/t$ and the 
scaling analysis must be performed in the asymptotic regime
$\eta(N) < \gamma E_{\text{opt}}$, the required system sizes
$N$ increases exponentially as $t/E_{\text{opt}}$ for 
$U/t \rightarrow 0$.  
Fortunately, it is not necessary to carry out this analysis
for the Hubbard model because the optical conductivity
of the weak-coupling field theory
is already in excellent agreement with the optical conductivity
of the lattice model for $U = 3t$.

\subsection{Strong-coupling limit \label{sec:strong}}

In this section I discuss the special case of a Mott-Peierls
insulator ($\Delta \neq 0, U > 0$) in the strong-coupling
limit $U/t \rightarrow \infty$, for which the shape of the
optical spectrum is known analytically.~\cite{florian2}
In this limit there is exactly one electron on each site in the ground 
state of the Peierls-Hubbard Hamiltonian~(\ref{hamiltonian}).
An optical excitation moves one electron from a site to another
and thus creates a double occupation (doublon) and an empty site
(holon). 
Therefore, the optical gap is of the order of $U$.
These elementary charge excitations are spinless bosons as in the 
Hubbard model.
The properties of the spin degrees of freedom are determined by an 
effective Heisenberg model with alternating exchange couplings $J_1 
\sim (t+\Delta/2)^2/U$ and $J_2 \sim (t-\Delta/2)^2/U$.
The spin gap $E_s$ vanishes in the limit $U \rightarrow \infty$. 
However, as there is a gap in the spin excitation spectrum
for any finite $U/t$ (see also next section), the structure of the spin 
ground state in the limit $U/t \rightarrow \infty$ is actually 
similar to that of a gapped state.    
For instance, the antiferromagnetic spin-spin correlations
decreases exponentially with increasing distance.
Thus, this strong-coupling limit of the Peierls-Hubbard model
is different from the two limiting cases discussed previously and 
from the general case presented in the next section.

\begin{figure}
\includegraphics[width=8cm]{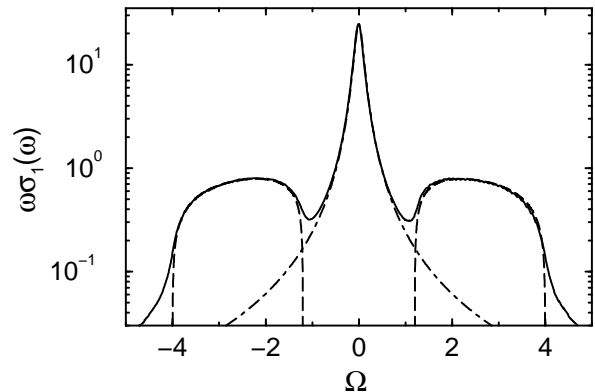}
\caption{ \label{fig4}
Reduced optical conductivity $C_{\eta}[\omega \sigma_1(\omega)]$
as a function of $\hbar \Omega = \hbar \omega - U$
in the strong-coupling limit $U \gg t$
for $\Delta=0.6t$. The solid line is the DDMRG result, 
Eq.~(\ref{Crsigma}), for
a 128-site lattice with a broadening $\eta/t=0.1$.
The dot-dashed line is a Lorentzian distribution of width $\eta/t=0.1$
centered at $\hbar \omega=U$.
The two dashed lines represent the analytical result 
(\ref{strongOC1}) for the continuum of an infinite system ($\eta=0$).
Note the logarithmic scale of the ordinate axis.
}
\end{figure}

In the thermodynamic limit the optical conductivity can be calculated 
analytically using some reasonable assumptions.~\cite{florian2}
If $0 < |\Delta| < 2t$, the spectrum  consists of two bands
for $2|\Delta| \leq |\hbar\omega -U| \leq 2t$
\begin{equation}
\sigma_1(\omega) = \frac{g_0 e^2 a}{8\hbar} 
\frac{\sqrt{[(\hbar \Omega)^2-(2\Delta)^2]
[(4t)^2 -(\hbar \Omega)^2]}}{\hbar \omega |\hbar\Omega|} \; ,
\label{strongOC1}
\end{equation}
where $\hbar \Omega = \hbar \omega - U$,
and a $\delta$-peak at $\hbar \omega = U$  
\begin{equation}
\sigma_1(\omega) = \frac{\pi g_{\pi} e^2 a t^2}{\hbar U}
\delta(\hbar \omega-U) 
\label{strongOC2}
\end{equation}
in the middle of the gap $4 |\Delta|$ separating the bands. 
For $\Delta \rightarrow 0$ one recovers the optical spectrum
of the Hubbard model in the strong coupling limit,
which consists of a single band and a $\delta$-peak in the middle
of this band.~\cite{eric,florian2}
The prefactors $g_0$ and $g_{\pi}$ are spin form factors given
by ground-state spin correlation functions.
They are functions of the effective exchange coupling ratio $J_2/J_1$ 
and thus of $\delta = |\Delta/2t|$. 
Assuming a dimer spin ground state (i.e., $J_1 > 0$ and $J_2=0$)
one obtains $g_0 = 9/4$ and 
\begin{equation}
g_{\pi} = \frac{1+32 \delta +62 \delta^2 + 32 \delta^3 + \delta^4}
{4 (1 + \delta)^2} \; .
\end{equation}
This result becomes exact in the dimer limit $|\Delta|=2t$, where 
$g_{\pi}=8$.
For $\Delta\rightarrow 0$ the dimer spin ground state does not give
the correct form factors 
because it is known exactly that $g_0 + g_{\pi} = 4 \ln(2)$
and it was found numerically that $g_{\pi}/g_0 \approx 10^{-2}$ 
(see Ref.~\onlinecite{eric}).

Figure~\ref{fig4} shows the reduced optical conductivity 
$C_{\eta}[\omega \sigma_1(\omega)]$ calculated using DDMRG on a 
128-site lattice for $\Delta=0.6t$ and $\eta/t=0.1$.
A logarithmic scale is used to make visible the weak bands on both sides
of the strong central peak.
In Fig.~\ref{fig4} one can recognize the spectral shape predicted by the
strong-coupling analysis.
To make a quantitative comparison, however, it is first necessary to 
determine $g_0$ and $g_{\pi}$ using the finite-size scaling analysis
of Sec.~\ref{sec:scaling}.
Here I use a size-dependent broadening $\eta(N)/t = 12.8/N$ as for the 
Hubbard model because the spectral width is also of the order of 
$8t$.
For $\hbar \omega = U-2t$ DDMRG results for
$C_{\eta}[\omega \sigma_1(\omega)]$ tend to 0.78 (in units of 
$e^2at/\hbar^2$) for $N \rightarrow \infty$.
Comparison with Eq.~(\ref{strongOC1}) then yields $g_0\approx 2.2$.
In Fig.~\ref{fig4} I also show the two bands~(\ref{strongOC1})
with this value of $g_0$ (without broadening). 
The agreement with the finite-system DDMRG spectrum is excellent. 
The small deviations visible close to the band edges are due to the 
different broadening used for the numerical result ($\eta/t=0.1$)
and for the analytical result ($\eta=0$). 
They vanish if the same broadening is used in both calculations.
Once more this confirms that a broadening satisfying 
Eq.~(\ref{etacondition}) hides most finite-size effects in this model
as already shown by other examples in Sec.~\ref{sec:peierls} of
this paper and in Refs.~\onlinecite{eric} and \onlinecite{fabian}.
In the DDMRG spectra $C_{\eta}[\omega \sigma_1(\omega)]$
the height of the central peak diverges as $2.42t/\eta$ 
(in units of $e^2at/\hbar^2$) for increasing $\eta(N)$ but its 
position does not change. 
This confirms that it corresponds to a $\delta$-peak
at $\hbar \omega =U$ and gives an estimate $g_{\pi} \approx 2.42$.
This $\delta$-peak broadened with a Lorentzian distribution of width
$\eta/t=0.1$ is also shown in Fig.~\ref{fig4}.
One sees that the agreement with the DDMRG result is perfect.
A similar finite-size scaling was performed to determine the form 
factor $g_{\pi}$ in the Hubbard model.~\cite{eric}

One notes the surprisingly good agreement between the form factors 
determined numerically with DDMRG 
($g_0 \approx 2.2$ and $g_{\pi} \approx 2.42$) 
and those obtained using the approximation of a dimer spin ground 
state ($g_0 = 2.25$ and $g_{\pi} \approx 2.52$).
For the  value $\Delta = 0.6t$ used in this example, the
ratio $J_2/J_1 \approx 0.29$ of the effective exchange coupling is 
already quite small and thus the dimer spin ground state is probably
a very good approximation of the actual spin ground state.

\subsection{Mott-Peierls insulator \label{sec:mottpeierls}}

\begin{figure}
\includegraphics[width=7cm]{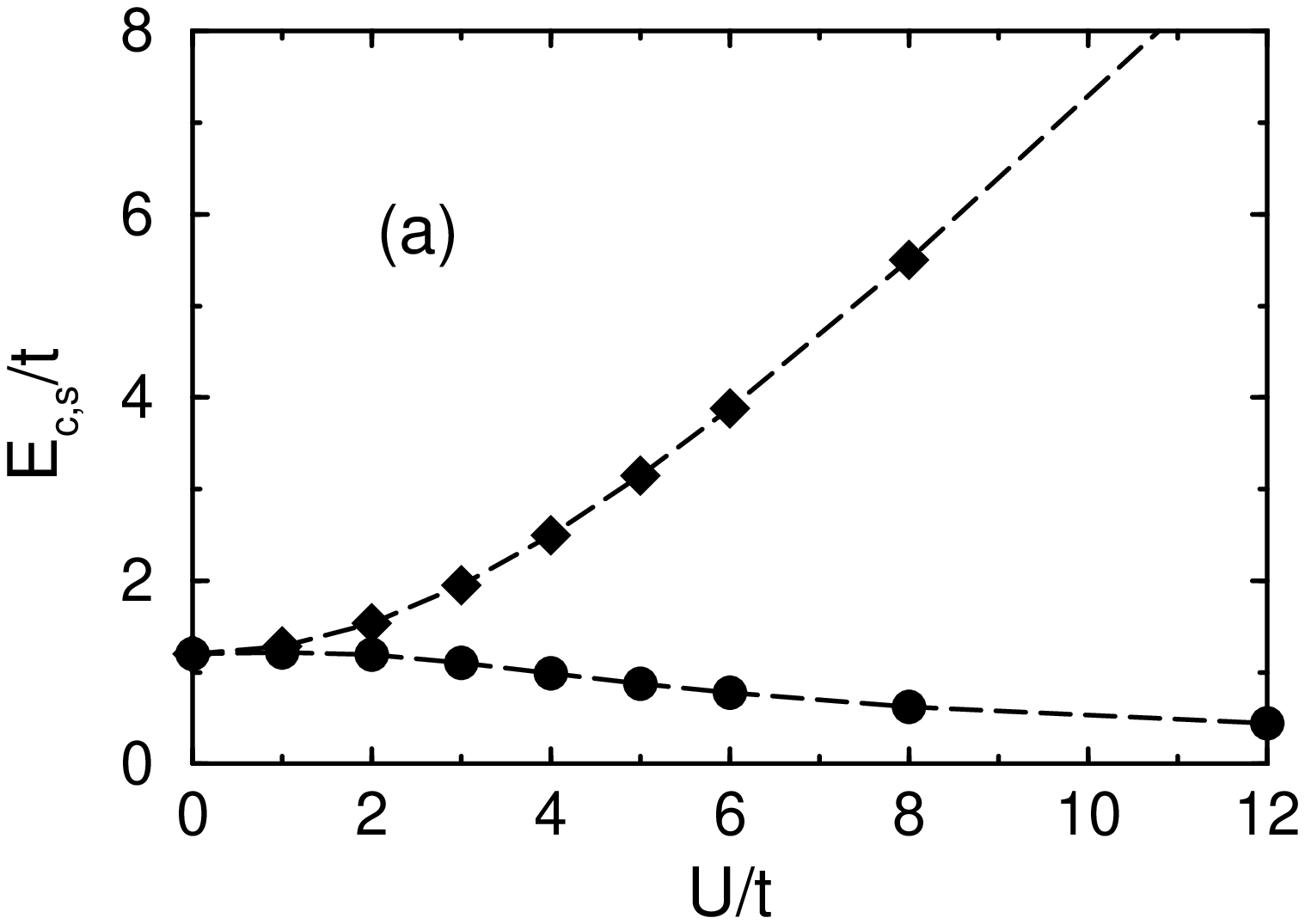}
\includegraphics[width=7cm]{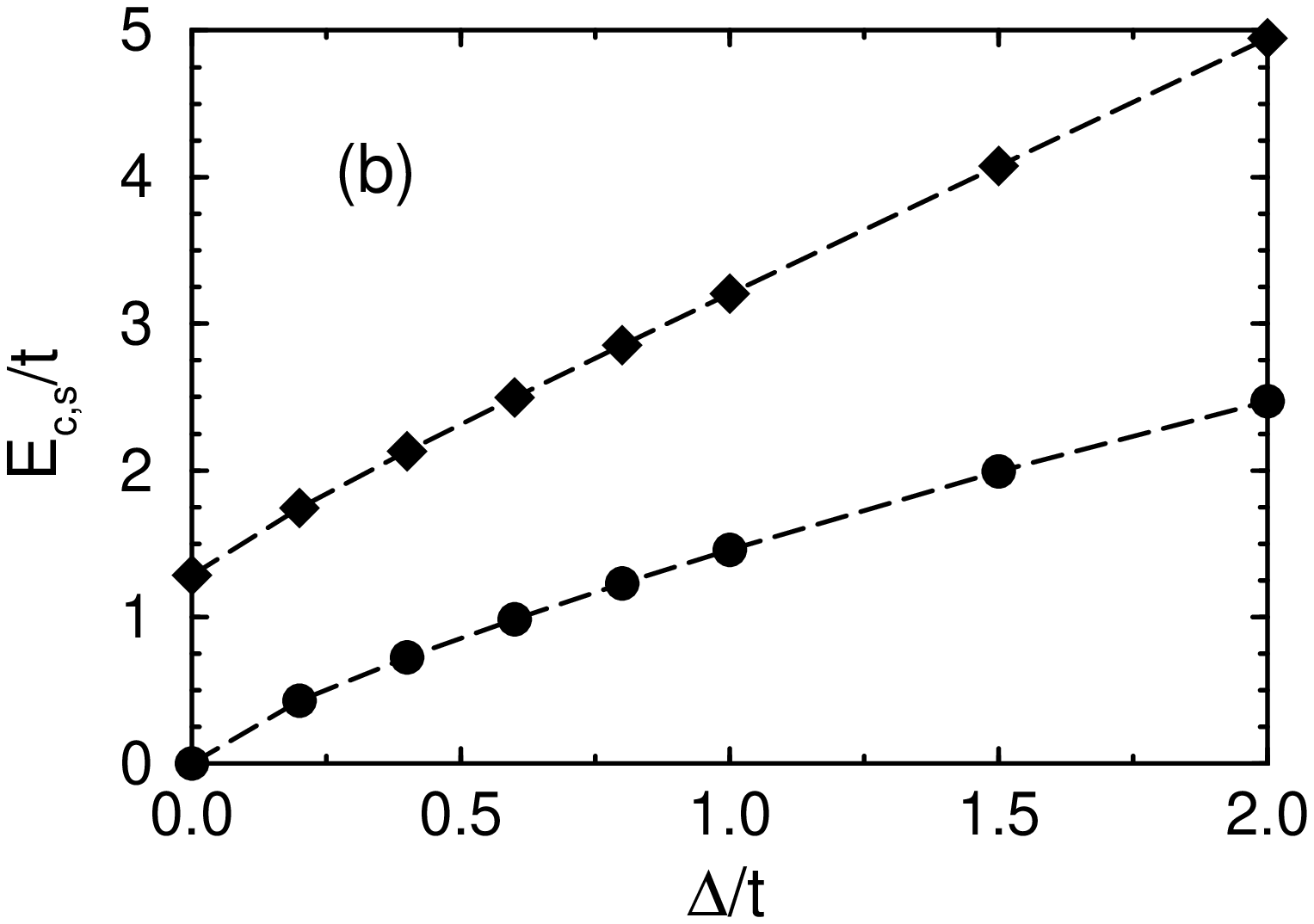}
\caption{ \label{fig5}
Charge (diamond) and spin (circle) gaps extrapolated to the 
thermodynamic limit. (a) As a function of $U$ for $\Delta=0.6t$. 
(b) As a function of $\Delta$ for $U=4t$.
}
\end{figure}

The optical conductivity of the Peierls-Hubbard model is not known
for general interaction parameters $U > 0 $ and $2t > |\Delta| > 0$.
In this regime the system is in a Mott-Peierls insulating 
phase~\cite{dionys}:
both a periodic lattice potential (i.e., the alternating hopping
terms) and electronic correlations contribute to the formation
of a charge gap $E_c > 0$ and there is a finite spin gap $E_s > 0$.
Numerical investigations of the charge and spin gaps and of static
correlation functions reveal no phase transition at finite $U$ and
intermediate $\Delta$  
(see also Ref.~\onlinecite{dionys} and references therein).
Thus the entire parameter space ($0 < U/t < \infty$,
$0 < |\Delta| < 2t$) belongs to a single Mott-Peierls insulating
phase.
Figure~\ref{fig5} shows charge and spin gaps as a function of $U$
and $\Delta$.
These gaps have been calculated on lattices with up to $N=400$ sites 
using DMRG and extrapolated to the thermodynamic limit. 
The charge gap of the Mott-Peierls insulator is always larger than the
gap of the Mott-Hubbard and Peierls insulators in the $\Delta=0$ and 
$U=0$ limits, respectively.
The spin gap is always smaller than the charge gap in the Mott-Peierls 
phase. 

In the thermodynamic limit the optical gap $E_{\text{opt}}$
is equal to the charge gap.
The nature of the optical excitations in the Mott-Peierls insulator
is not well understood.
Despite the obvious difference between charge and spin 
excitation energies, $E_s < E_c$, it is not even known if there is
a spin-charge separation for single-particle excitations.
Optical excitations could consist of a pair of fermionic quasi-particles
with opposite spins $\pm \sigma$ and opposite charges $\pm e$ as in a 
Peierls insulator (Sec.~\ref{sec:peierls}).
They could as well be made of two spinless bosonic excitations
carrying opposite charges $\pm e$ as in the Mott-Hubbard insulator
(Sec.~\ref{sec:hubbard}) and in the strong-coupling limit 
(Sec.~\ref{sec:strong}).

The investigation of spin and charge gaps and static correlation
functions clearly shows that the three special cases described in the
previous sections are singular limits of the Peierls-Hubbard model.
Unsurprisingly, I have found that the optical conductivity in the
Mott-Peierls phase is unlike the simple spectrum
observed in these limits.
[All optical spectra presented in this section have been calculated
using DDMRG and the finite-size scaling analysis has always been 
performed using a size-dependent broadening $\eta(N)/t=12.8/N$.]

\begin{figure}
\includegraphics[width=8cm]{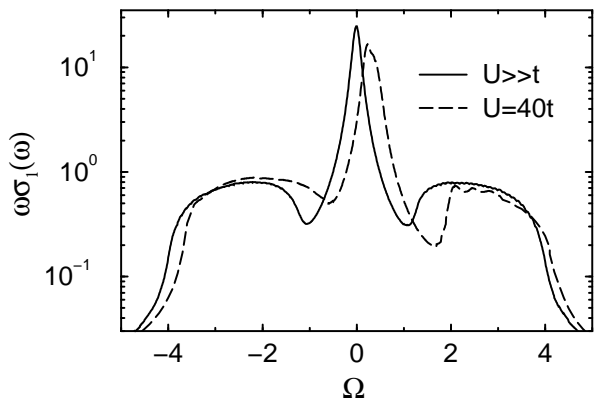}
\caption{ \label{fig6}
Reduced optical conductivity $C_{\eta}[\omega \sigma_1(\omega)]$ 
calculated with DDMRG [see Eq.~(\ref{Crsigma})]
on a 128-site lattice ($\eta/t=0.1$) 
in the strong-coupling regime for $\Delta=0.6t$ 
as a function of $\hbar \Omega= \hbar \omega-U$.
Note the logarithmic scale of the ordinate axis.
}
\end{figure}

For large but finite $U$ the optical spectrum consists of
three bands: a narrow band with a strong singularity around 
$\hbar \omega = U$ and one weak band on each side of this central peak.
The singularity seems to be made of two very close power-law divergences
which merge to form the single isolated $\delta$-peak (\ref{strongOC2})
in the $U/t \rightarrow \infty$ limit.
The optical spectrum starts as $\sqrt{\hbar\omega-E_{\text{opt}}}$
at the lower band edge $E_{\text{opt}}$ for all $|\Delta| < 2t$.
Figure~\ref{fig6} shows the reduced optical conductivity 
$C_{\eta}[\omega \sigma_1(\omega)]$ for $U=40t$ and $\Delta=0.6t$.
The spectrum looks very similar to the spectrum for 
$U/t \rightarrow \infty$, which is again shown in this figure.
A finite-size-scaling analysis shows however that the strong central
peak is not a $\delta$-function for $U=40t$ but a narrow band
with at least one singularity diverging as $\eta^{-0.8}$.
The spectra in Fig.~\ref{fig6} are made of three bands: the gaps
between the bands appear as local minima on both sides of the central 
peak because of the relatively large broadening used ($\eta/t=0.1$).
The finite-size scaling analysis confirms the existence of these gaps.
For decreasing parameters $U$ or $|\Delta|$ first the lower gap,
then the upper gap close.
Therefore, the number of bands in the optical spectrum of the
Mott-Peierls insulator is not constant but depend on the interaction
parameters $U$ and $\Delta$.

\begin{figure}
\includegraphics[width=8cm]{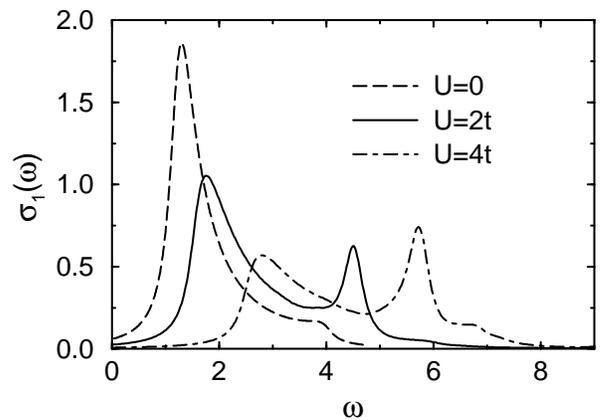}
\caption{ \label{fig7}
Optical conductivity $\sigma_1(\omega)$ 
calculated with DDMRG on a 64-site lattice ($\eta/t=0.2$)
for $\Delta=0.6t$ and several values of $U$. 
}
\end{figure}

The evolution of the optical conductivity as a function of $U$
is very interesting.
For decreasing $U/t$ one observes that the central peak breaks into
two peaks appearing as local maxima in the broadened spectrum
of finite-size systems. 
The first peak (at the lowest energy) takes over most of the 
optical weight of the central peak. 
Its weight decreases progressively with decreasing $U/t$ but
remains strong even for small $U$.
In Fig.~\ref{fig7}, it is clearly visible (at $\hbar \omega > 4t$)
even for $U=2t$ (with $\Delta=0.6t$). 
This peak corresponds to a power-law divergence within a band 
with an exponent  that tends to $-1/2$ for $U \rightarrow 0$.
The peak position moves to lower energy as $U$ decreases and reaches
$\hbar \omega = 4t$ for $U=0$.
Therefore, the central peak observed at strong coupling $U \gg t$
corresponds to the upper square-root divergence in the Peierls 
insulator spectrum~(\ref{peierlsOC}).
(In Fig.~\ref{fig7} this divergence is barely visible as a local 
maximum at $\hbar \omega \approx 4t$ because of the relatively
large broadening $\eta/t =0.2$ used.)
The second peak has very little optical weight (it is not visible
in Fig.~\ref{fig7}) and I have not been able to determine its structure.

For $U \gg t$ the optical weight is distributed symmetrically around
the central peak, as seen in Fig.~\ref{fig6}.
As $U$ decreases, there is a progressive transfer of optical weight 
from high frequency (above the central peak) to low frequency 
(below the central peak), see Fig.~\ref{fig7}.
The high-frequency spectrum becomes very weak for small $U/t$ but it 
completely disappears only at $U=0$.
It is difficult to determine the spectral width for general 
parameters because the optical conductivity
is very weak and vanishes smoothly at high frequency.
I estimate that the width of the spectrum lies between $4t$ and $8t$ 
for $U > 0$.
The smallest width is reached for large $\Delta$ and small
$U$, the largest for small $\Delta$ and large $U$.
The low-frequency spectrum becomes stronger as $U$ diminishes.
The local maximum below the central peak (seen in Fig.~\ref{fig6})
progressively rises, moves closer to the lower band edge, and 
transforms into a strong narrow peak, visible in the spectra
shown in Fig.~\ref{fig7} (at $\hbar \omega < 4t$).
For small enough $U/t$ this low-energy peak contains more optical 
weight than the central peak.
For $U \rightarrow 0$ the low energy peak becomes the
square-root divergence of the Peierls insulator 
spectrum~(\ref{peierlsOC}) at the band threshold $E_{\text{opt}}$.

For $U>0$, however, my results suggest that the optical spectrum 
always vanishes smoothly at the optical gap.
I think that the low-energy optical spectrum of the Peierls-Hubbard 
model at weak coupling has a qualitative behavior similar to that of 
the Hubbard model:     
as $\hbar \omega - E_{\text{opt}}$ decreases, 
$\sigma_1(\omega)$ first appears to diverge as 
$(\hbar \omega - E_{\text{opt}})^{-1/2}$, then goes through a maximum
just above the optical gap $E_{\text{opt}}$, and vanishes smoothly 
for $\hbar \omega \rightarrow E_{\text{opt}}$.
For large enough $U$ it is possible to carry out a finite-size scaling 
analysis similar to the one performs for the Hubbard model 
(see Sec.~\ref{sec:hubbard}).
Thus it is possible to check that the low-energy spectrum maximum is 
finite and 
lies at a higher energy than the optical gap, and to show explicitly 
that $\sigma_1(\omega) \sim \sqrt{\hbar \omega - E_{\text{opt}}}$ 
for $\hbar \omega - E_{\text{opt}} \rightarrow 0^+$.

\begin{figure}
\includegraphics[width=8cm]{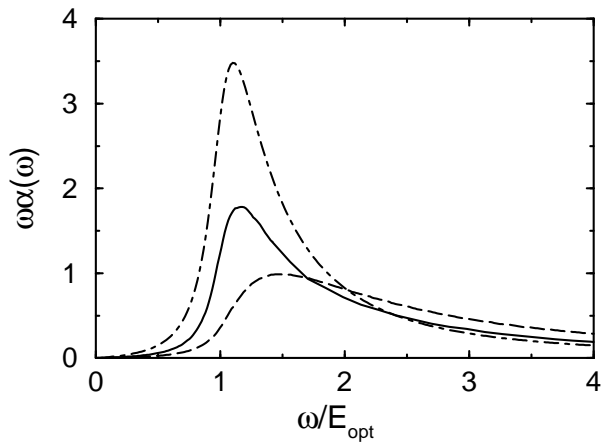}
\caption{ \label{fig8}
Optical conductivity $\omega C_{\eta}[\alpha(\omega)]$ 
calculated with DDMRG [see Eq.~(\ref{Calpha})]
on a 128-site lattice ($\eta/t=0.1$)
in the small gap regime: Mott-Hubbard insulator with $U=3t$ (dashed),
Peierls insulator with $\Delta=0.3t$ (dot-dashed), and 
Peierls-Hubbard insulator with $U=2.3t$ and $\Delta=0.15t$ (solid).
The optical gaps are $E_{\text{opt}}= 0.631t$, $0.6t$, and $0.704t$,
respectively.
}
\end{figure}

For smaller $U$ it becomes increasingly difficult to distinguish
a smooth spectrum with a truncated divergence from a true divergence. 
For instance, Fig.~\ref{fig8} shows the low-frequency optical 
conductivity $\omega C_{\eta}[\sigma_1(\omega)/\omega]$ for $U=2.3t$ 
and $\Delta=0.15t$ with a broadening $\eta/t=0.1$ ($N =128$ sites).
For comparison, I also show the spectra in two limits discussed 
previously, the Mott-Hubbard insulator 
[$\sigma_1(\omega) \sim \sqrt{\hbar \omega - E_{\text{opt}}}\ $]
and the Peierls insulator [$\sigma_1(\omega) \sim 1/ \sqrt{\hbar 
\omega - E_{\text{opt}}} \ $], with similar optical gaps 
($E_{\text{opt}}/t=0.6-0.7$) and the same broadening $\eta$.
Clearly, the Mott-Peierls insulator spectrum looks like an 
intermediate case between the spectra observed in both limiting cases.
The position of the maximum in the Mott-Peierls insulator spectrum
tends to $0.78t$ for $\eta(N) \rightarrow 0$ while the charge gap 
(and optical gap) equals $0.704t$ in the thermodynamic limit.
Certainly, there is no divergence in the low-energy optical spectrum. 
However, the maximum seems to diverge as $1/\sqrt{\eta}$ even
for the smallest broadening I have used ($\eta/t = 0.05$).
Thus this spectrum seems to be qualitatively similar to  
a function such as Eq.~(\ref{typicalOC}),
but the maximum is so close to the optical gap that
broadenings $\eta$ significantly smaller than $0.05t$ (i.e., system 
sizes much larger than $N =256$) would be necessary to reach the
asymptotic regime as discussed in Sec.~\ref{sec:scaling}. 
For the same reason it is not possible to determine how the
spectrum vanishes for $\hbar\omega - E_{\text{opt}} \rightarrow 0^+$
in such a case

In the Hubbard model it is possible to confirm the absence of
a singularity and the square-root vanishing at the band
threshold even 
if the optical gap is as small as $E_{\text{opt}}=0.4t$, because we 
know the optical spectrum of an infinite system for $E_{\text{opt}} 
\rightarrow 0$ from field theory.~\cite{eric}
The field theory approach does not only apply to the Hubbard model, but
more generally, gives the low-energy optical spectrum of one-dimensional
Mott insulators with small Mott gaps.~\cite{controzzi,fabian}
The different spectral functions depend only on an interaction parameter
$\beta^2 \leq 1$.
In addition, the optical gap $E_{\text{opt}}>0$ and a normalization 
constant set the frequency scale and the conductivity scale.
For $1/2 < \beta^2 \leq 1$ these optical spectra described 
truncated square-root divergence with a square-root vanishing at the 
band threshold as in Eq.~(\ref{typicalOC}).
For $\beta^2=1$ (Hubbard model) the spectrum has the shape shown in 
Fig.~\ref{fig3} with a maximum at $1.24 E_{\text{opt}}$.
As $\beta^2$ decreases the peak becomes sharper and the maximum 
moves closer to the band edge.
For $\beta^2=1/2$ the optical spectrum is similar to that of 
a Peierls band insulator with a square-root divergence at the band
threshold. 

\begin{figure}
\includegraphics[width=8cm]{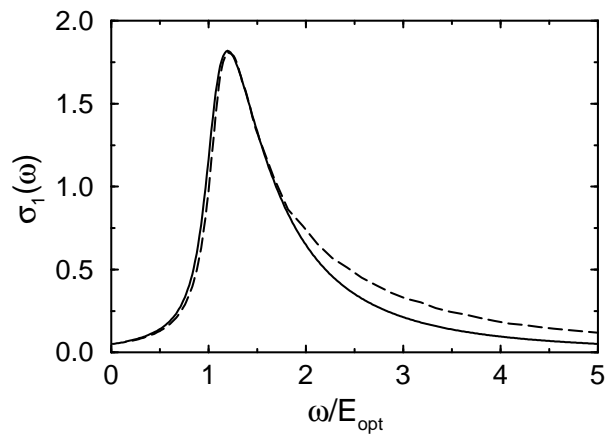}
\caption{ \label{fig9}
Optical conductivity $\sigma_1(\omega)$ of a Mott-Peierls
insulator 
calculated with DDMRG on a 128-site lattice ($\eta/t=0.1$)
for $U=2.3t$ and $\Delta=0.15t$ (dashed) and the fitted
field-theoretical spectrum for Mott insulators with $\beta^2=0.58$ 
and the same broadening $\eta/t$ (solid).
}
\end{figure}

Therefore, this field theory~\cite{fabian,controzzi}
can describe the optical spectrum in both limiting cases 
(Mott-Hubbard and Peierls insulators) of the Peierls-Hubbard model 
in the small gap regime,
and the field-theoretical spectrum evolves continuously from one limit 
to the other with $\beta^2$ going from 1 to 1/2.
Using $1 \geq \beta^2 \geq 1/2$, the optical gap $E_{\text{opt}}$, and 
the normalization constant as fit parameters, I have compared the 
low-energy optical conductivity calculated with DDMRG for small gaps 
($E_{\text{opt}} < 0.71t$) to field-theoretical spectra with similar
broadening as explained in Sec.~\ref{sec:scaling}.
For instance, I show in Fig.~\ref{fig9} the DDMRG spectrum
for the lattice model~(\ref{hamiltonian}) with $U=2.3t$ and 
$\Delta=0.15t$ and the fitted field-theoretical spectrum with 
$\beta^2=0.58$.
Both spectra agrees up to $\hbar \omega = 1.2t \approx 1.7 
E_{\text{opt}}$. 
Generally, I have found that the optical spectrum of a Mott-Peierls 
insulator can be fitted by a field-theoretical spectrum with 
$\beta^2 > 1/2$ over a finite frequency range, from $\omega =0$ to a 
frequency $\omega$ which lies between the position of the low-energy 
maximum and $2E_{\text{opt}}/\hbar$.   
(Naturally, for $U\rightarrow 0$ the best fit is always obtained with 
$\beta^2=1/2$.)
Therefore, I think that for any $U>0$ (and $0 \leq |\Delta| < 2t$)
the optical spectrum vanishes as $\sqrt{\hbar \omega - E_{\text{opt}}}$
for $\hbar \omega - E_{\text{opt}} \rightarrow 0^+$.

Note that I do not assume that the field-theoretical calculations in
Refs.~\onlinecite{fabian} and~\onlinecite{controzzi} are also valid
for the Peierls-Hubbard model with general interaction parameters.  
Actually, there are visible discrepancies starting at rather low energy 
between field theory and DDMRG results for the lattice model, as shown
in Fig.~\ref{fig9}. 
The agreement between field theory and DDMRG results in the region of
the band threshold
simply means that the optical spectra in Mott-Peierls insulators
and in one-dimensional Mott insulators have similar shapes just above 
the optical gap.  

\begin{figure}
\includegraphics[width=7cm]{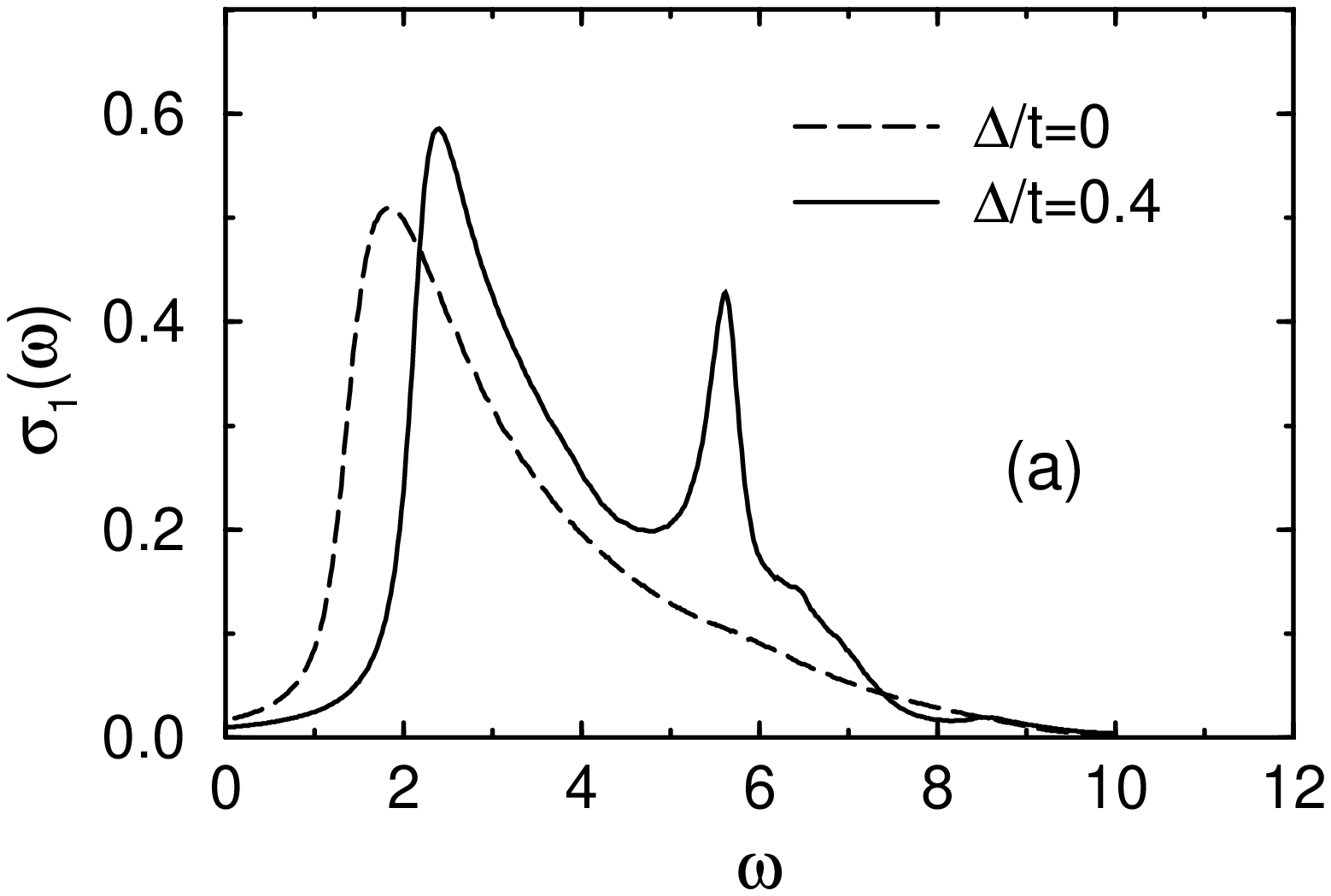}
\includegraphics[width=7cm]{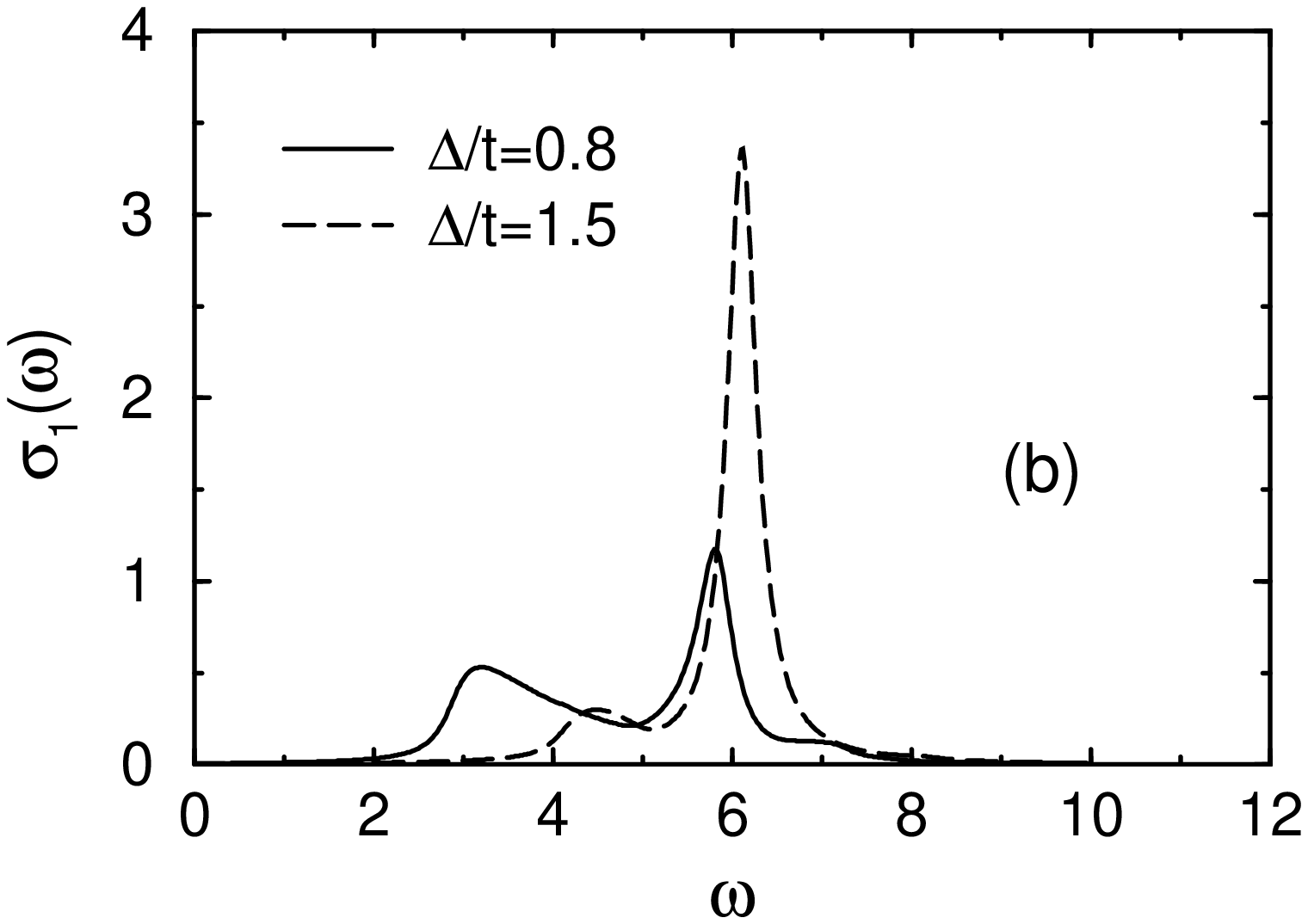}
\caption{ \label{fig10}
Optical conductivity $\sigma_1(\omega)$ 
calculated with DDMRG on a 64-site lattice ($\eta/t=0.2$)
for $U=4t$ and several values of $\Delta$. 
(a) For $r(\Delta) > 0.5$. 
(b) For $r(\Delta) < 0.5$.
}
\end{figure}

Finally, it is interesting to examine the evolution of the optical 
spectrum from weak to strong bond alternation for fixed $U$.
It has been shown in Ref.~\onlinecite{eric} 
(see also Sec.~\ref{sec:hubbard}) that for $\Delta=0$ (Hubbard 
model) the spectrum consists of a single band with a maximum close to 
the lower band edge and a tiny peak in the center (at least for 
$U \geq 4t$).
If $|\Delta|$ increases one observes in Fig.~\ref{fig10}(a)
that the maximum moves closer to the optical gap and corresponds to a 
sharper peak.
The optical spectrum still starts as 
$\sqrt{\hbar \omega - E_{\text{opt}}}$ at the 
band threshold as discussed above.
The central peak, which is too weak to be seen in Fig.~\ref{fig10}(a)
for $\Delta=0$, becomes rapidly stronger as $|\Delta|$ increases and 
is clearly visible for $\Delta=0.4t$.  
As discussed previously this peak becomes a $\delta$-function in the
strong-coupling limit $U\gg t$ and corresponds to the upper square-root
divergence of the Peierls insulator if $U$ vanishes. 
For moderate $|\Delta|$ the ratio between the hopping integrals 
$r(\Delta) = (t-|\Delta|/2)/(t+|\Delta|/2)$ is not too small and 
the optical weight is mostly concentrated below the central peak.
If this ratio becomes small, however, the central peak becomes
the spectrum dominant feature, see Fig.~\ref{fig10}(b). 
The proportion of the optical weight which is in the central peak 
increases as $1-r^2(\Delta)$ for $r(\Delta) \rightarrow 0$.
A finite-size scaling analysis confirms however that this peak is 
not a $\delta$-function but is still a power-law divergence within an 
excitation band. 
Only in the dimer limit $|\Delta|=2t$ [$r(\Delta)=0$] the optical 
spectrum is made of a single $\delta$-peak, which corresponds to the
excitation of Frenkel excitons localized on a dimer. 

In summary, the optical spectrum of a Mott-Peierls insulator
consists of one or more bands with a total spectral width ranging
from $4t$ to $8t$.
The distinctive features of the spectrum are a square-root
vanishing of $\sigma_1(\omega)$ at the lower band edge
and a peak due to a power-law singularity around the middle of the 
spectrum.
For strong couplings [$U \gg t$ and $r(|\Delta|) \ll 1$]
most of the optical weight is in the central peak,
while for weak couplings [$U \ll 4t$ and $r(|\Delta|) > 1/2$]
it is mostly concentrated in a narrow peak just above the optical gap. 
In the limit of a vanishing gap ($U \rightarrow 0$ and $\Delta 
\rightarrow 0)$ this narrow peak becomes a Drude peak at $\omega=0$.
For intermediate couplings most of the optical weight is distributed
over a broad frequency range between the optical gap and the central 
peak 
and $\sigma_1(\omega)$ goes through a maximum in this range.

The central peak always appears at an energy larger than the bare band
width $4t$.
For parameters which are realistic for conjugated polymers,\cite{dionys}
most of the optical weight lies below this peak.
Therefore, I think that it is not possible to observe such
a structure in the optical spectrum of conjugated polymers, because 
it occurs at a too high energy ($>$ 10~eV) and its intensity
is too weak. 

The optical spectrum of Mott-Peierls insulators is unlike that
of Mott-Hubbard and Peierls insulators.
However, most of its main features are found in the strong-coupling 
limit investigated in Ref.~\onlinecite{florian2} and discussed
in Sec.~\ref{sec:strong}.
This suggests that the optical excitations of a Mott-Peierls insulator
could be made of a pair a spinless bosonic excitations with opposite 
charges as in the strong-coupling limit (and in a Mott-Hubbard 
insulator).
Nevertheless, understanding the nature of the system elementary 
excitations requires the study of additional dynamical properties, 
such as the one-particle Green's functions.
The DDMRG method will enable us to carry out this further investigation.

\section{Conclusion \label{sec:conclusion}}

In this paper I have presented a dynamical DMRG method which allows one
to calculate the optical conductivity of one-dimensional correlated
electron systems on large lattices with great accuracy. 
The DDMRG approach to the calculation of dynamical properties
is essentially an application of the standard DMRG algorithm for
ground-state calculations. 
Therefore, both methods have the same advantages but also the same 
limitations.
In particular, DDMRG will directly benefit from recent and future 
improvements of DMRG such as the use of additional symmetries. 

With DDMRG it is possible to calculate the dynamical response of 
correlated systems with hundreds of sites and particles.
Relative errors of the order of $10^{-4}$ can be achieved for the 
optical spectrum of finite systems with $N \sim 10^2$ sites.
Using a finite-size scaling analysis based on a size-dependent 
broadening of the discrete finite-system spectra, one can then 
calculate a dynamical spectrum in the thermodynamic limit with a 
resolution of the order of 1\% of the spectral width and even 
investigate singularities in a continuum.

The DDMRG approach can be used for various dynamical quantities, such as
dynamical spin-spin correlation functions or single-particle Green's 
functions.
It can also be applied to other lattice quantum many-body models,
in higher dimension, including spin or boson degrees of freedom, or 
long-range interactions.
The correction vector DMRG method has been used to calculate 
non-linear dynamic response functions, such as third-order
dynamical polarizabilities.~\cite{pati}
Similarly, the variational principle presented in 
Sec.~\ref{sec:principle} can be generalized to dynamical
correlation functions describing these non-linear responses. 
Thus I believe that it is possible to develop an efficient
DDMRG method for calculating these quantities.
A limitation of the DDMRG approach is the restriction to zero 
temperature.
It would be desirable to extend the variational principle and the DDMRG
approach to finite-temperature dynamical properties. 

The computational resources used by the DDMRG method are relatively
modest.
For instance, all calculations presented in this paper were carried out
on workstations with a single 500 MHz Alpha processor and required less
than 1 GByte of memory. 
It would be easy and very efficient to run DDMRG on a parallel computer
as calculations for different frequencies are almost independent.
This would permit one to investigate much larger or more complicated 
systems than in this work.

In summary, the DDMRG method and the finite-size scaling
technique for dynamical spectra appear extremely accurate
and versatile. They provide a powerful
approach for investigating the dynamical properties 
in low-dimensional lattice quantum many-body systems.

Finally, it should be kept in mind that the variational principle
for dynamical correlation functions and their related excited states
is completely independent from the DMRG method.
Therefore, it is possible to combine this principle with other
variational methods to calculate dynamical properties.
For instance, one could build a trial wavefunction
$|\psi(\{\lambda_i\})\rangle = R(\{\lambda_i\}) |\psi_0\rangle$,
where $R(\{\lambda_i\})$ is an operator depending on a few parameters
$\lambda_i$, such that the calculation of $W(\{\lambda_i\})=
W_{A,\eta}(\omega,\psi(\{\lambda_i\}))$ reduces to the evaluation 
of ground-state correlation functions.
Then the minimization of $W(\{\lambda_i\})$ with respect to the 
variational parameters $\{\lambda_i\}$ would give a lower bound and an 
approximate value of the dynamical spectrum $I_{A}(\omega+ i \eta)$.

\begin{acknowledgments}
I am grateful to F.~Essler and F.~Gebhard for many stimulating 
conversations. 
I also acknowledge useful discussions with T.D.~K\"{u}hner 
and S.R. White about the correction vector DMRG method.  
\end{acknowledgments}


\end{document}